\documentclass[journal]{IEEEtran}
\ifCLASSINFOpdf
\else
\fi
\ifCLASSOPTIONcompsoc
 \usepackage[caption=false,font=normalsize,labelfont=sf,textfont=sf]{subfig}
\else
 \usepackage[caption=false,font=footnotesize]{subfig}
\fi
\def\Nt{{N_{\mathrm{t}}}}
\def\Nr{{N_{\mathrm{r}}}}
\def\Nf{{N_{\mathrm{f}}}}
\def\Ntx{{N_{\mathrm{t},x}}}
\def\Nrx{{N_{\mathrm{r},x}}}
\def\Nty{{N_{\mathrm{t},y}}}
\def\Nry{{N_{\mathrm{r},y}}}
\def\nt{{n_{\mathrm{t}}}}
\def\nr{{n_{\mathrm{r}}}}
\def\nf{{n_{\mathrm{f}}}}
\def\ntx{{n_{\mathrm{t},x}}}
\def\nrx{{n_{\mathrm{r},x}}}
\def\nty{{n_{\mathrm{t},y}}}
\def\nry{{n_{\mathrm{r},y}}}

\def\df{{d_{\mathrm f}}}
\def\dtx{{d_{{\mathrm t},x}}}
\def\drx{{d_{{\mathrm r},x}}}

\def\dty{{d_{{\mathrm t},y}}}
\def\dry{{d_{{\mathrm r},y}}}

\def\one{{\boldsymbol{1}}}

\def\bell{{\boldsymbol{\ell}}}

\def\th{{\tilde{h}}}
\def\tD{{\tilde{D}}}

\def\lambdac{\lambda_{\mathrm{c}}}

\def\bb0{{\mathbb{0}}}

\def\bb{{\boldsymbol{b}}}

\def\bee{{\boldsymbol{e}}}

\def\bk{{\boldsymbol{k}}}

\def\bmm{{\boldsymbol{m}}}
\def\bn{{\boldsymbol{n}}}

\def\br{{\boldsymbol{r}}}

\def\bt{{\boldsymbol{t}}}

\def\bz{{\boldsymbol{z}}}
\def\b0{{\boldsymbol{0}}}

\def\bI{{\boldsymbol{I}}}

\def\bN{{\boldsymbol{N}}}

\def\bR{{\boldsymbol{R}}}

\def\b{{\mathrm{b}}}

\def\r0{{\mathbf{0}}}

\def\bbC{{\mathbb{C}}}

\def\bbE{{\mathbb{E}}}

\def\bbN{{\mathbb{N}}}

\def\bbR{{\mathbb{R}}}

\def\bbZ{{\mathbb{Z}}}

\def\cD{\mathcal{D}}

\def\cL{\mathcal{L}}
\def\cM{\mathcal{M}}
\def\cN{\mathcal{N}}
\def\cO{\mathcal{O}}

\def\sfD{\mathsf{D}}

\def\btheta{\bm \theta}

\def\bdelta{\bm \delta}

\def\bomega{\bm \omega}

\def\bsf0{{\bm{\mathsf{0}}}}

\def\N0{{N_{\mathrm{0}}}}

\def\vec{\mathrm{vec}}

\def\fc{f_{\mathrm{c}}}

\def\bsf{{\boldsymbol{s}_\mathrm{f}}}

\def\Nf {{N_\mathrm{f}}}

\newcommand{\data}{\mathbf{y}}

\newcommand{\be}{\begin{equation}}
\newcommand{\ee}{\end{equation}}
\newcommand{\bal}{\begin{align}}
\newcommand{\eal}{\end{align}}

\def\SNR    {{\mathsf{SNR}}}

\def\fc {f_{\mathrm{c}}}
\def\lambdac {\lambda_{\mathrm{c}}}

\usepackage{graphicx}
\usepackage{amsmath}
\usepackage{amsthm}
\usepackage{mathtools}
\usepackage{epsfig}
\usepackage{float}
\usepackage{lipsum}
\usepackage{stfloats}
\usepackage{array}
\usepackage{amssymb}
\usepackage{cite}
\usepackage{url}
\usepackage{algorithm}
\usepackage{algpseudocode}
\usepackage{multirow}
\usepackage{fancyh}
\usepackage{flushend}
\usepackage{upgreek}
\usepackage{dsfont}
\usepackage{gensymb}
\usepackage{booktabs}
\usepackage{bm}
\usepackage[shortlabels]{enumitem}
\usepackage{threeparttable}
\usepackage{physics}
\usepackage{nicematrix}

\usepackage{multicol, blindtext}
\usepackage{mathtools, cuted}

\usepackage{color}
\usepackage{cancel}

\usepackage{tikz}
\usepackage{pgfmath, pgfplots}
\usepackage{pgfplotstable}
\usetikzlibrary{3d}
\pgfplotsset{table/search path={data}}

\usepackage{pifont}
\newcommand{\cmark}{\ding{51}}%
\newcommand{\xmark}{\ding{55}}%

\theoremstyle{remark}

\theoremstyle{remark}

\allowdisplaybreaks
\DeclareMathOperator*{\argmax}{argmax} 
\DeclareMathOperator*{\argmin}{argmin} 

\normalsize
\begin{document}
%
\title{Near-Field Channel Estimation Via \\ Wavefront Parameterization 
\thanks{
Heedong Do and A. Lozano are with Univ. Pompeu Fabra, 08018 Barcelona (e-mail: angel.lozano@upf.edu). Their work is supported by
the Departament de Recerca i Universitats de la Generalitat de Catalunya
and by the Maria de Maeztu Units of Excellence Programme CEX2021-001195-M funded by MICIU/AEI/10.13039/501100011033. 
Namyoon Lee is with POSTECH, Korea 37673 (email: nylee@postech.ac.kr). His work is supported by the National Research Foundation of Korea (NRF) grants (No. RS-2023-00208552, No. RS-2022-NR070834), funded by the Korea government (MSIT).
This article was presented in part at the Asilomar Conference on Signals, Systems, and Computers 2024 \cite{do2024estimation}.
}
}
\author{\IEEEauthorblockN{Heedong~Do},
 {\it Member,~IEEE},
\and
\IEEEauthorblockN{Namyoon~Lee},
{\it Senior Member,~IEEE},
\and\\
\IEEEauthorblockN{Angel~Lozano},
{\it Fellow,~IEEE}
\vspace{-8mm}
}

\maketitle

\begin{abstract}
This paper deals with the estimation of multiantenna channels in the line-of-sight conditions that are prevalent in the near field. By expressing the curved wavefront as a polynomial via a power series expansion of a sphere, the estimation of the channel over the array can be formulated as a multidimensional polynomial phase estimation problem. The application of a newly developed polynomial phase estimator, able of handling arbitrary dimensions and polynomial degrees, yields a superior tradeoff between channel estimation accuracy and complexity.

\end{abstract}



%
\IEEEpeerreviewmaketitle

\section{Introduction}


With the push to higher frequencies and, especially, to larger array apertures, a growing share of wireless transmissions is anticipated to take place in near-field conditions,
with wavefronts whose curvature is noticeable 
\cite{do2021terahertz, zhang20236g, ramezani2023exploiting}.
An enticing attribute of near-field propagation is that spatial multiplexing becomes possible in line-of-sight (LOS) or via a clean reflection \cite{pizzo2023wide}, sidestepping the dependence on multipath richness that far-field channels are contingent on \cite{driessen1999capacity,jiang2005spherical,bohagen2005construction, sarris2007design, bohagen2007design, bohagen2007optimal, bohagen2009spherical, torkildson2011indoor,song2015spatial,do2020capacity,do2021reconfigurable,pizzo2022landau,ding2022degrees,do2023line}.
This motivates the growing interest in the estimation of near-field LOS channels \cite{zhang2023near2,liu2024near,yang2025near}.

Although one could ignore the high degree of structure in an LOS channel and directly estimate the channel entries connecting every transmit with every receive antenna \cite{996869}, vastly more accurate estimates can be produced by recognizing that these entries---however many---are functions of a few geometric parameters. Moreover, some (or all) of this increase in estimation accuracy can then be traded for a drastically reduced pilot overhead.

An abundance of model-based parameter estimators have been proposed (see Table~\ref{table:priorArt}, top section), often capable of handling multipath components in addition to the LOS link, and of capitalizing on observations at multiple frequencies.
A considerable number of works have also thrown 
learning at the
problem (see Table~\ref{table:priorArt}, bottom section). 
On the one hand, this is appealing because parameter estimation boils down to the extraction of hidden quantities. On the other hand, though, learning methods offer no performance guarantees and their accuracies are thus exceedingly difficult to compare.
By the same token, their complexities cannot be contrasted, as that would require putting their performance on a common footing.

Model-based parametric channel estimators do offer performance guarantees, and they
are indeed vastly superior to their antenna-domain counterparts, albeit at a computational cost that explodes with the number of geometric parameters. 
As detailed later, this number ranges from two (for a single antenna at one end of the link and a linear array at the other) to six (for planar arrays at both ends), with parametric estimators having hitherto been considered for up to four parameters. This leaves out the arguably most important cases, namely a planar array at one end and a linear array at the other, or planar arrays at both ends.



\begin{table*}
\setlength{\tabcolsep}{3pt}
\addtolength{\leftskip} {-2cm}
\addtolength{\rightskip} {-2cm}
\centering
\begin{threeparttable}
\caption{Prior Art on Near-Field Parametric Channel Estimation \\ Top Section: Model-Based Methods, Including the Present Paper. Bottom Section: Learning Methods }
\label{table:priorArt}
\begin{tabular}{c | c c | c c c | c | c} 
\toprule
& \multicolumn{2}{c|}{Array} & \multicolumn{3}{c|}{Channel model} & Observation via & 
\\
Reference & Topologies & Orientation & Wavefront & Multipath & Multiple frequencies & hybrid arrays\tnote{*} & Parameters
\\
\midrule
\cite{le2019massive} & linear + single & - & spherical/parabolic & \cmark & \xmark & \xmark & 1 distance, 1 angle
\\
\cite{zhu2025adaptive} & linear + single & - & spherical & \cmark & \cmark & \cmark & 1 distance, 1 angle
\\
\cite{huang2020one} & linear + single & - & parabolic & \cmark & \xmark & \xmark & 2 polynomial coefficients
\\
\cite{cui2022channel} & linear + single & - & spherical & \cmark & \cmark & \cmark & 1 distance, 1 angle
\\
\cite{zhang2024near} & linear + single & - & spherical & \cmark & \xmark & \xmark & 1 distance, 1 angle
\\
\cite{hussain2024near} & linear + single & - & spherical & \xmark & \cmark & \cmark & 1 distance, 1 angle
\\
\cite{das2024multipath} & linear + single & - & parabolic & \cmark & \cmark & \xmark & 2 polynomial coefficients
\\
\cite{wang2025mbpd} & linear + single & - & spherical & \cmark & \cmark & \cmark & 1 distance, 1 angle
\\
\cite{xu2024near} & linear + single & - & parabolic & \cmark & \xmark & \xmark & 1 distance, 1 angle
\\
\cite{11434893} & linear + single & - & parabolic & \cmark & \xmark & \cmark & 1 distance, 1 angle
\\
\cite{chung2025efficient} & planar + single & - & parabolic & \cmark & \xmark & \xmark & 1 distance, 2 angles
\\
\cite{kosasih2023parametric} & planar + single & - & spherical & \xmark & \xmark & \xmark & 1 distance, 2 angles
\\
\cite{lu2024near} & planar + single & - & spherical & \cmark & \xmark & \xmark & 1 distance, 2 angles
\\
\cite{li2024hybrid} & planar + single & - & spherical & \cmark & \xmark & \xmark & 1 distance, 2 angles
\\
\cite{yuan2025scalable} & planar + single & - & piecewise planar & \xmark & \xmark & \xmark & 1 distance, 2 angles
\\
\cite{ghermezcheshmeh2023parametric} & planar + single & - & piecewise planar & \cmark & \xmark &\xmark & 1 distance, 2 angles
\\
\cite{chen2026near} & planar + single & - & parabolic & \cmark & \cmark & \cmark & 1 distance, 2 angles
\\
\cite{lu2023near} & linear + linear & 2D & spherical & \cmark & \xmark &  \cmark & 1 distance, 2 angles
\\
\cite{guo2025channel} & linear + linear & 2D & spherical & \cmark & \xmark &  \cmark & 1 distance, 2 angles
\\
\cite{song2025line} & linear + linear & any & ellipsoidal & \xmark & \xmark &  \cmark & 1 distance, 4 angles
\\
\cite{shi2025double} & linear + linear & any & parabolic & \cmark  & \xmark &\cmark & 5 polynomial coefficients
\\
\textbf{This paper} & planar + planar & any & polynomial & \xmark & \cmark & \xmark & any set of polynomial coefficients
\\
\midrule
\cite{zhang2023near} & linear + single & - & parabolic & \cmark & \xmark &\cmark & 1 distance, 1 angle \\
\cite{lei2024channel} & linear + single & - & spherical & \cmark & \xmark & \xmark & 1 distance, 1 angle\\
\cite{jang2024neural} & linear + single & - & spherical & \cmark & \cmark & \cmark & 1 distance, 1 angle\\
\cite{wang2025near} & linear + single & - & spherical & \cmark & \cmark & \xmark & 1 distance, 1 angle \\
\cite{zheng2024model} & linear + single & - & spherical & \cmark & \cmark & \xmark & 1 distance, 1 angle \\
\cite{yu2024bayes} & linear + single & - & spherical & \cmark\tnote{$\dagger$} & \xmark & \cmark & 1 distance, 1 angle \\
\cite{tao2026deep} & linear + single & - & parabolic & \cmark & \cmark & \cmark & 1 distance, 1 angle \\
\cite{yu2023adaptive} & planar + single & - & spherical & \cmark & \cmark
& \cmark & 1 distance, 2 angles \\
\cite{fang2025near} & linear + linear & 2D & spherical & \cmark & \xmark & \xmark & 1 distance, 2 angles \\
\cite{chen2021hybrid} & planar + planar & parallel & piecewise planar & \cmark & \xmark & \cmark & 1 distance, 4 angles \\
\cite{lin2025deep} & planar + planar & fixed Tx & spherical & \cmark & \cmark & \cmark & 1 distance, 2 angles
\\
\bottomrule
\end{tabular}
\begin{tablenotes} \footnotesize
\item[*] With fully digital arrays, the channel matrix itself can be observed. 
With hybrid arrays, instead, a compressed version---the product of the analog combining matrix, the channel, and the analog precoding matrix---is observed. Since digital arrays are a special case of hybrid arrays in which the analog stages are immaterial, estimation methods suitable for hybrid arrays can be applied to digital arrays, but not vice versa.
\item[$\dagger$] This work differs from the rest in that it adopts a correlated Rayleigh fading model. The near-field effect is incorporated in the correlation matrix.
\end{tablenotes}
\end{threeparttable}
\end{table*}

\begin{figure}
    \centering
    \begin{tikzpicture}
      \draw (0,0) ellipse (1.25 and 0.75);
      \fill[fill=white] (-1.25,0.2) rectangle (1.25,0.8);
      \node[align=center] at (0,0.5) {LOS channels\\[-3pt]\footnotesize(6 geometric parameters)};
      \draw (0,0) ellipse (2.5 and 1.5);
      \fill[fill=white] (-1.8,1) rectangle (1.8,1.6);
      \node[align=center] at (0,1.3) {Polynomial phase channels\\[-3pt]\footnotesize($|\cM|$ polynomial coefficients)};
      \draw (0,0) ellipse (3.75 and 2.25);
      \fill[fill=white] (-1.4,1.9) rectangle (1.4,2.5);
      \node[align=center] at (0,2.1) {Arbitrary channels\\[-3pt]\footnotesize($2 \Nt\Nr$ real entries)};
    \end{tikzpicture}
    \caption{Venn diagram illustrating the inclusion of sets of channels. Here, $\Nt$ and $\Nr$ are the number of transmit and receive antennas while $\cM$ is the set of polynomial degrees (see Sec. \ref{sec:wavefront_parameterization}.) }
    \label{fig:venn_diagram}
\end{figure}
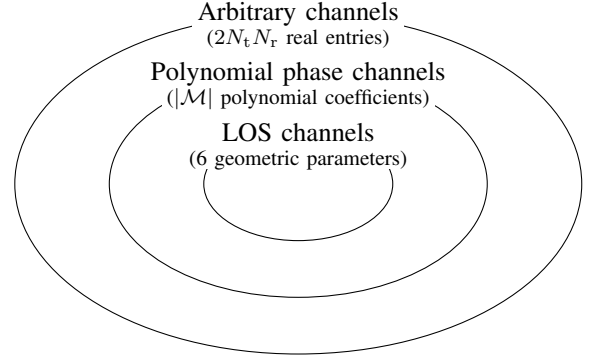

Against this backdrop, this paper propounds a middle way (see Fig. \ref{fig:venn_diagram}) to tackle the problem of estimating near-field LOS channels.
Rather than estimating the channel entries directly, or the hidden geometric parameters that determine those entries, this middle way seeks to estimate the curvature and disposition of the wavefronts. While nominally spherical, the wavefronts can be approximated to any desired accuracy by a polynomial function \cite{do2023parabolic}. (The far-field approximation is a limiting case in which this polynomial is an affine function.)
Once the polynomial coefficients necessary for the desired accuracy have been estimated from noisy pilot observations, at either one or multiple frequencies,
the channel entries can be straightforwardly reconstructed with strict performance and complexity guarantees.

As a starting point to explore this middle way, this paper considers moderately large arrays over which the channel amplitude variations can be neglected, such that only the phase reflects the near-field behavior. Subsequent work will tackle the extension to extremely large arrays over which the amplitudes also experience significant variations. 

The polynomial expansion of the phase variations yields a channel whose phase is polynomial in the antenna and frequency indices. This invites a polynomial phase estimator, and in particular a recently developed such estimator that can handle multivariate polynomials of
arbitrary degrees \cite{do2025multidimensional}. This estimator is efficient at high SNR, attaining the Cramer-Rao bound (CRB), yet its complexity is orders of magnitude below that of existing geometric parameter estimators. This offers the promise of a performance commensurate with that of these estimators, but at a drastically more affordable complexity, even with planar arrays at  both ends of the link.

The paper is organized as follows. After Sec. \ref{sec:models} presents the LOS channel model, Sec. \ref{sec:parametric_channel_estimation} introduces the notion of parametric estimation, and the two subsequent sections dwell on the geometric and wavefront parameterizations, respectively. Then, Sec. \ref{sec:polynomial_phase_estimation} imports the multidimensional polynomial phase estimation method that is applied to the latter parameterization. A performance evaluation is provided in Sec. \ref{sec:numerical_evaluation}, followed by a summary and discussion in Sec. \ref{sec:conclusion}.

\subsection{Notation}
\label{sec:notation}

The set of nonnegative integers is denoted by $\bbN_0$ while the first $N$ nonnegative integers are compactly denoted by
\begin{align}
\label{UoM}
    [N]\equiv \{0,1,\ldots,N-1\}.
\end{align}

The generalized binomial coefficient is, for $n\in\bbZ$, denoted by \cite{graham1994concrete}
\begin{align}
    {n \choose k} \equiv \begin{cases}
    \frac{n(n-1)\cdots(n-k+1)}{k!} & \;\; \text{integer }k\geq0\\
    0 & \;\; \text{integer }k<0 
    \end{cases} .
    \label{EOS}
\end{align}

Multidimensional counterparts to \eqref{UoM} are also used. Specifically, for
\begin{align}
    &\bN=(N_0,\ldots,N_{\sfD-1})\in\bbZ^\sfD \qquad\! \bn=(n_0,\ldots,n_{\sfD-1})\in\bbZ^\sfD\nonumber\\
    & \bmm=(m_0,\ldots,m_{\sfD-1})\in\bbZ^\sfD \qquad\! \bk=(k_0,\ldots,k_{\sfD-1})\in\bbN_0^\sfD \nonumber\\
    &\br=(r_0,\ldots,r_{\sfD-1})\in\bbR^\sfD \nonumber
\end{align}
let us define $[\bN]\equiv [N_0]\times \cdots \times [N_{\sfD-1}]$ with $\times$ the Cartesian product of sets, as well as
\begin{align}
    \br^\bk &\equiv r_0^{k_0}\cdots r_{\sfD-1}^{k_{\sfD-1}} 
    & \qquad {\bn \choose \bmm}&\equiv {n_0 \choose m_0}\ldots{n_{\sfD-1} \choose m_{\sfD-1}} .
\end{align}

\begin{figure}
    \centering
    \begin{tikzpicture}[>=stealth]
    \begin{scope}[scale = 1.5]
        \draw[fill=gray!50, draw=black, shift={(0.2, 0.7)}]
              (0, 0) to[out=30, in=140] (1.5, -0.2) to [out=60, in=160]
              (5, 0.5) to[out=130, in=60]
              cycle;
        \shade[thin, left color=gray!10, right color=gray!50, draw=black,
              shift={(0.2, 0.7)}]
              (0, 0) to[out=20, in=140] (3, -0.8) to [out=60, in=190] (5, 0.5)
                to[out=130, in=60] cycle;
        \filldraw (1.8,1.7) circle (1pt) node[left]{$h$};
        \filldraw (2.3,1.9) circle (1pt) node[right]{$\hat{h}$};
        \filldraw (2,3) circle (1pt) node[right]{$y = h + w_{\bbC}$};
        \draw[line width=1pt, dotted, ->] 
            (2,3) -- (2.27,1.94) 
            node[midway, right, align=left] {};
    \end{scope}
    \end{tikzpicture}
    \caption{The MLE projects the observation on the channel manifold, denoising it. In the small-error regime, the manifold becomes flat and only the noise components on the $M$-dimensional tangent plane survive, with power $\frac{M}{2 \, \SNR}$.}
    \label{fig:channel_manifold}
\end{figure}
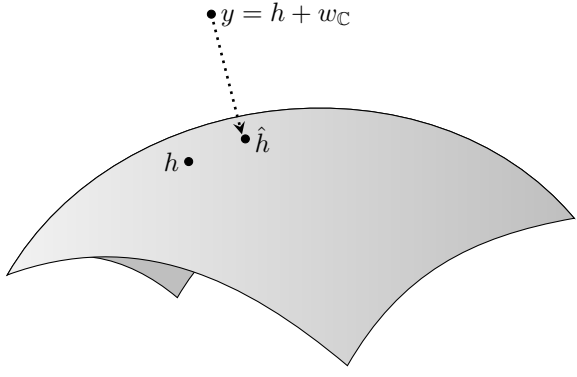

\section{Channel and Observation Models}
\label{sec:models}

\subsection{Channel Model}

Consider
$\Nf$ equispaced frequencies centered on $\fc$ with normalized interval $\df$, meaning
\begin{equation}
f_{\nf} = \fc \left( 1 + \df \left(\nf- \frac{\Nf-1}{2} \right)\right) \qquad   \nf \in [\Nf].
\end{equation}
Transmitter and receiver feature uniform planar arrays (UPAs) whose dimensionalities are $\Ntx\times \Nty$ and $\Nrx\times \Nry$, with $\Nt=\Ntx\Nty$ and $\Nr=\Nrx\Nry$. The antenna indices are $(\ntx,\nty) \in [\Ntx]\times[\Nty]$ and $(\nrx,\nry)\in[\Nrx]\times[\Nry]$, with $\nt \in [\Nt]$ and $\nr \in [\Nr]$ sometimes used for the sake of conciseness; the explicit mapping rules of these bijections are immaterial.
The antenna spacings along the planar dimensions are $\dtx$ and $\dty$ at the transmitter, and $\drx$ and $\dry$ at the receiver.  Linear arrays and single antennas are special cases in which some of the dimensions collapse into singletons.

The channel is either LOS, or subject to a specular reflection \cite{pizzo2023wide}. The focus is on near-field situations, with the far field subsumed as a special case.
Under the proviso that the amplitude variations across the entries are negligible, the channel connecting the $\nt$th transmit antenna with the $\nr$th receive antenna at frequency $f_{\nf}$ can be normalized into \cite{do2024estimation} 
\begin{align}
    h(\bn) = \frac{D}{D_{\nr,\nt}}\!\exp\! \bigg( \! -j\frac{2\pi }{\lambdac}  D_{\nr,\nt}\bigg(\!1 + \df \bigg(\!\nf- \frac{\Nf-1}{2}\!\bigg)\!\bigg)\!\bigg)\label{channel}
\end{align}
for $\bn\in[\bN]$,
where $D$ is the distance between the array centers, $D_{\nr,\nt}$ is the distance between the antennas, and $\lambdac$ is the wavelength corresponding to $\fc$, while
\begin{align}
    \bn &= (\nrx,\nry,\ntx,\nty,\nf)\\
    \bN &= (\Nrx,\Nry,\Ntx,\Nty,\Nf).
\end{align}


\subsection{Observation Model}

By means of pilot transmissions, orthogonal across transmit antennas and frequencies, a noisy version of the channel entries can be procured, namely
\begin{align}
    y(\bn) = h(\bn) + w_\bbC(\bn), \label{observation_model}
\end{align}
where $w_{\bbC}(\bn) \overset{\mathrm{iid}}{\sim} \cN_{\bbC}(0,\frac{1}{\SNR})$ is white Gaussian noise. 

For a channel estimate $\hat{h}(\bn)$, the per-entry mean-square error (MSE) is
\begin{align}
   \text{MSE} = 
   \frac{1}{|[\bN]|}
   \sum_{\bn\in[\bN]}  \! \bbE \! \left[ | \hat{h}(\bn)-h(\bn) |^2\right] \label{per_entry_mse}.
\end{align}
The standard least-squares (LS) channel estimate, which is also the maximum-likelihood estimate (MLE), boils down to the observation itself, i.e., $\hat{h}(\bn) = y(\bn)$. The per-entry MSE of the LS estimate is $\frac{1}{\SNR}$ \cite[Sec. 2.7]{foundations2018}.

\begin{figure}
    \centering
    \begin{tikzpicture}[>=stealth]
    \begin{scope}[scale = 1.5]
        \draw[fill=gray!50, draw=black, shift={(0.2, 0.7)}]
              (0, 0) to[out=30, in=140] (1.5, -0.2) to [out=60, in=160]
              (5, 0.5) to[out=130, in=60]
              cycle;
        \shade[thin, left color=gray!10, right color=gray!50, draw=black,
              shift={(0.2, 0.7)}]
              (0, 0) to[out=20, in=140] (3, -0.8) to [out=60, in=190] (5, 0.5)
                to[out=130, in=60] cycle;
        \draw[draw=black, line width = 1pt]
        (0.5, 0.79) to[out=60, in=190] (1.8, 1.7) to [out=10, in=160] (3, 0.7) to [out=340, in=210] (2.9,1.8) to [out=30, in=130] (4.45,1);
        \filldraw (1.8,1.7) circle (1pt) node[below]{$h$};
        \filldraw (2,3) circle (1pt) node[right]{$y = h + w_{\bbC}$};
    \end{scope}
    \end{tikzpicture}
    \caption{Even if the channel is known to lie on a lower-dimensional manifold (the one-dimensional curve), optimization over a higher-dimensional manifold (the two-dimensional surface) may be computationally advantageous---at the expense of a higher number of parameters.
    }
    \label{fig:different_parameterizations}
\end{figure}
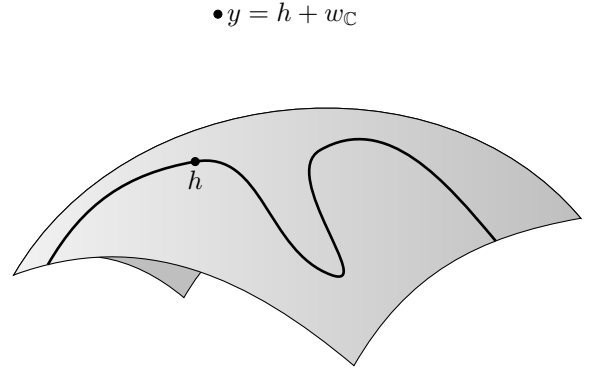

\section{Parametric Channel Estimation}
\label{sec:parametric_channel_estimation}

When the channel (LOS or multipath) is induced by a few parameters, vectorized into $\btheta \in  \bbR^M$, one can consider estimating $\btheta$ from $y(\bn)$ in \eqref{observation_model}.
The ensuing estimate $\hat{\btheta}$ can then be mapped to a channel estimate via $\hat{h}(\bn) \equiv h(\bn;\hat{\btheta})$.





Provided that the parameters are not redundant, the constraint set, i.e., the set
of all possible channels induced by $\btheta$ at each frequency, forms an $M$-dimensional manifold within the $2 \Nt \Nr$-dimensional space in which the channel nominally resides. 
For an unbiased\footnote{At high SNR, the bias of nice estimators---including the constrained MLE \cite{moore2008maximum}---tends to vanish and the unbiasedness premise becomes fully pertinent.} estimator, the application of the constrained CRB \cite[Example 2]{do2026cramer}
(see also \cite[Prop. 2]{lyu2026cramer})
indicates that the per-entry MSE in \eqref{per_entry_mse} is bounded below by
\begin{align}
    \frac{M}{2 \, |[\bN]|}\cdot\frac{1}{\SNR} = \frac{M}{2 \, \Nt \Nr \Nf}\cdot\frac{1}{\SNR}. \label{per_entry_crb}
\end{align}
This does not depend on the detailed structure of the manifold, because the CRB is a local quantity; the manifold locally resembles an $M$-dimensional subspace, and the power of the noise components thereon is $\frac{M}{2 \, \SNR}$ (see Fig. \ref{fig:channel_manifold}). 

The LS estimate, $\hat{h}(\bn) = y(\bn)$, can be seen as an extreme case of parametric channel estimation where the channel is parameterized by its $M=2 \Nt \Nr \Nf $ real-valued entries. 
%
%
%
More generally, channels can be parameterized in various manners, each entailing its pros and cons. The efficacy of a parameterization can be assessed by figures of merit, chiefly:
\begin{enumerate}
    \item \textit{Accuracy.} The parameterization should faithfully model the channel of interest. The high-SNR MSE performance is essentially curtailed by model mismatches.
    \item \textit{Number of parameters.} As per \eqref{per_entry_crb}, with a faithful model the number of parameters dictates the high-SNR MSE.
    \item \textit{Availability of computationally efficient algorithms.} Some parameterizations have a computational advantage over others \cite{hayes2023sinusoidal}.
\end{enumerate}
Trade-offs arise among the above, for instance between the second and the third merits as illustrated in Fig. \ref{fig:different_parameterizations}.
In fact, this trade-off between number of parameters and computational efficiency is one of the thrusts behind the wavefront parameterization propounded in this paper.

The next two sections contrast respective approaches to parameterize near-field LOS channels, namely (i) the geometric parameterization in Sec. \ref{sec:geometric_parameterization}, and (ii) the wavefront parameterization in Sec. \ref{sec:wavefront_parameterization}.
The latter drastically simplifies the estimation procedure, at the cost of being slightly redundant
in that the number of parameters (i.e., polynomial coefficients)
is higher than the number strictly needed to describe the channel.



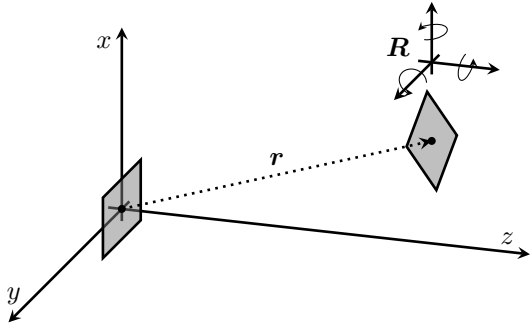
\begin{figure}
    \centering
    \begin{tikzpicture}[
        scale=1, 
        >=stealth, 
        x={(0 cm,0.8 cm)}, 
        y={(-0.5 cm,-0.5 cm)}, 
        z={(0.9 cm,-0.1 cm)}
    ]
        \draw[line width=1pt, ->] 
            (-0.2,0,0) -- (3,0,0) 
            node[above left, yshift = -0.4 cm] {$x$};
        \draw[line width=1pt, ->]
            (0,-0.2,0) -- (0,3,0)
            node[above left, xshift = 0.3 cm, yshift = 0.1 cm] {$y$};
        \draw[line width=1pt, ->]
            (0,0,-0.2) -- (0,0,6)
            node[above, xshift = -0.3 cm] {$z$};

        \draw[line width=1pt, fill opacity = 0.5, fill = gray, rotate around x = 0, rotate around y = 0, rotate around z = 0] 
            (0.5,0.5,0) -- (-0.5,0.5,0) -- (-0.5,-0.5,0) -- (0.5,-0.5,0) -- cycle;
        \fill[] (0,0,0) circle (1.5pt);
            
        \begin{scope}[shift = {(1,-1,4)}]
            \draw[line width=1pt, fill opacity = 0.5, fill = gray, scale = 0.5, rotate around x = 55, rotate around y = -10, rotate around z = 15]
            (1,1,0) -- (-1,1,0) -- (-1,-1,0) -- (1,-1,0) -- cycle;
            \fill[] (0,0,0) circle (1.5pt);

            \begin{scope}[shift = {(1.3,0,0)}]
                \node[] at (0.2,0,-0.5) {$\bR$};
                \draw[line width=1pt,->] 
                    (-0.2,0,0) -- (1,0,0) 
                    node[above left, yshift = -0.4 cm] {};
                \draw[line width=1pt,->]
                    (0,-0.2,0) -- (0,1,0)
                    node[above left, xshift = 0.3 cm, yshift = 0.1 cm] {};
                \draw[line width=1pt,->]
                    (0,0,-0.2) -- (0,0,1)
                    node[above, xshift = -0.3 cm] {};
                \begin{scope}[canvas is yz plane at x=0.5]
                    \draw[->] (0.2, 0) arc (0:270:0.2) node[midway, left=5pt] {};
                \end{scope}
                \begin{scope}[canvas is zx plane at y=0.5]
                    \draw[->] (0.2, 0) arc (0:270:0.2) node[midway, below=5pt] {};
                \end{scope}
                \begin{scope}[canvas is xy plane at z=0.5]
                    \draw[->] (0.2, 0) arc (0:270:0.2) node[midway, right=5pt] {};
                \end{scope}
            \end{scope}

        \end{scope}

        \draw[line width=1pt, ->, dotted]
            (0,0,0) -- node[midway, above] {$\br$} (1,-1,4);
    \end{tikzpicture} 
    \caption{Two planar arrays and the parameters describing their relative geometry.}
    \label{fig:translation_and_orientation}
\end{figure}

\section{Geometric Channel Parameterization}
\label{sec:geometric_parameterization}

As shown in Fig. \ref{fig:translation_and_orientation}, the coordinate system is chosen so that the transmit array is on the $xy$-plane with its center at the origin.
The location of the $\nt$th transmit antenna is 
\begin{align}
    \bt_\nt = 
    \begin{bmatrix}
        \dtx \big( \ntx-\frac{\Ntx-1}{2} \big) \\ \dty \big(\nty-\frac{\Nty-1}{2} \big) \\ 0
    \end{bmatrix}
\end{align}
and that of the $\nr$th receive antenna can be expressed as
\begin{align}
    \br_\nr = \br + \bR
    \begin{bmatrix}
        \drx \big( \nrx-\frac{\Nrx-1}{2} \big) \\ \dry \big( \nry-\frac{\Nry-1}{2} \big) \\ 0
    \end{bmatrix} ,
\end{align}
where $\bR\in\bbR^{3\times 3}$ is a rotation matrix.
As the set of rotation matrices forms a three-dimensional manifold, the total number of geometric parameters is six for planar arrays; three translations (described by $\br$) and three rotations (described by $\bR$).
For linear arrays or a single antenna, the number of parameters is further reduced as indicated in Table \ref{table:parameter_count}. 

An alternative geometric parameterization \cite{bohagen2007design, bohagen2007optimal, bohagen2009spherical} is detailed in App. \ref{app:another_parameterization}; while the parameters are different (more rotations and fewer translations), their numbers are identical to those in Table \ref{table:parameter_count}.

\begin{table}
\setlength{\tabcolsep}{4pt}
\addtolength{\leftskip} {-2cm}
\addtolength{\rightskip} {-2cm}
\centering
\begin{threeparttable}
\caption{Number of parameters required to describe the relative geometry of two arrays}
\label{table:parameter_count}
\begin{tabular}{c c | c c c} 
\toprule
& & \multicolumn{3}{c}{\# parameters}\\
Tx array & Rx array & Translation & Rotation & Total\\
\midrule
linear & single & 2 & 0 & 2 \\
planar & single & 3 & 0 & 3\\
linear & linear & 2 & 2 & 4\\
planar & linear & 3 & 2 & 5\\
planar & planar & 3 & 3 & 6\\
\bottomrule
\end{tabular}
\end{threeparttable}
\end{table}

\subsection{Constrained Parameterization}

To curb the explosion in computational complexity,
in some prior art a constrained geometry is assumed 
to tacitly shrink the number of parameters.
In \cite{lu2023near, guo2025channel, fang2025near}, for instance, two linear arrays are coplanar, demanding three geometric parameters instead of four. In \cite{chen2021hybrid}, two planar arrays are parallel, requiring three geometric parameters instead of six.
In \cite{lin2025deep}, the center of the receive array is on the $z$-axis, calling for four parameters instead of six. (Conversely, in \cite{song2025line}, the relative geometry between two linear arrays is parameterized with one distance and four angles, which is redundant.) 

\subsection{Infeasibility of Brute-Force Search}
\label{sec:infeasibility}

\begin{figure*}
    \centering
    \subfloat[$(2\times 1)\times(1\times 1)$]{
        \begin{tikzpicture}
            \pgfplotstableread[col sep=comma]{data/trajectories_2x1_1x1_cost.csv}{\data}
            \begin{axis}[
                width=5.9cm, height=3.5cm,
                xlabel={Iteration},
                ylabel={Cost [dB]},
                xlabel style={yshift=1mm},
                ylabel style={yshift=-3mm},
                xtick={0, 100, ..., 500},
                xmin=0, xmax=500,
                ymin=-35, ymax=5,
                ytick={-100, -90, ..., 100},
                grid=both,
                legend pos=north east,
                legend cell align={left},
                tick label style={font=\footnotesize},
                label style={font=\small},
                legend style={font=\small}
            ]
            \addplot[red, opacity=0.4] table [x=Iteration, y = Best_1_Cost_dB] {\data};
            \addlegendentry{Random}
            \pgfplotsinvokeforeach{2, 4, 8, 16, 32, 64, 128, 256, 512, 1024}{
                \addplot[red, opacity=0.4, forget plot] table [x=Iteration, y=Best_#1_Cost_dB] {\data};
            }
            \addplot[line width=1.5pt, blue] table [x=Iteration, y=Proxy_Cost_dB] {\data};
            \addlegendentry{Genie}
            \end{axis}
        \end{tikzpicture}
    }
    \subfloat[$(8\times 1)\times(1\times 1)$]{
        \begin{tikzpicture}
            \pgfplotstableread[col sep=comma]{data/trajectories_8x1_1x1_cost.csv}{\data}
            \begin{axis}[
                width=5.9cm, height=3.5cm,
                xlabel={Iteration},
                ylabel={Cost [dB]},
                xlabel style={yshift=1mm},
                ylabel style={yshift=-3mm},
                xtick={0, 100, ..., 500},
                xmin=0, xmax=500,
                ymin=-13, ymax=3,
                ytick={-100, -95, ..., 100},
                grid=both,
                legend pos=north east,
                legend cell align={left},
                tick label style={font=\footnotesize},
                label style={font=\small},
                legend style={font=\small}
            ]
            \addplot[red, opacity=0.4] table [x=Iteration, y = Best_1_Cost_dB] {\data};
            \pgfplotsinvokeforeach{2, 4, 8, 16, 32, 64, 128, 256, 512, 1024}{
                \addplot[red, opacity=0.4, forget plot] table [x=Iteration, y=Best_#1_Cost_dB] {\data};
            }
            \addplot[line width=1.5pt, blue] table [x=Iteration, y=Proxy_Cost_dB] {\data};
            \end{axis}
        \end{tikzpicture}
    }
    \subfloat[$(32\times 1)\times(1\times 1)$]{
        \begin{tikzpicture}
            \pgfplotstableread[col sep=comma]{data/trajectories_32x1_1x1_cost.csv}{\data}
            \begin{axis}[
                width=5.9cm, height=3.5cm,
                xlabel={Iteration},
                ylabel={Cost [dB]},
                xlabel style={yshift=1mm},
                ylabel style={yshift=-3mm},
                xtick={0, 100, ..., 500},
                xmin=0, xmax=500,
                ymin=-13, ymax=3,
                ytick={-100, -95, ..., 100},
                grid=both,
                legend pos=north east,
                legend cell align={left},
                tick label style={font=\footnotesize},
                label style={font=\small},
                legend style={font=\small}
            ]
            \addplot[red, opacity=0.4] table [x=Iteration, y = Best_1_Cost_dB] {\data};
            \pgfplotsinvokeforeach{2, 4, 8, 16, 32, 64, 128, 256, 512, 1024}{
                \addplot[red, opacity=0.4, forget plot] table [x=Iteration, y=Best_#1_Cost_dB] {\data};
            }
            \addplot[line width=1.5pt, blue] table [x=Iteration, y=Proxy_Cost_dB] {\data};
            \end{axis}
        \end{tikzpicture}
    }\\
    \subfloat[$(2\times 1)\times(2\times 1)$]{
        \begin{tikzpicture}
            \pgfplotstableread[col sep=comma]{data/trajectories_2x1_2x1_cost.csv}{\data}
            \begin{axis}[
                width=5.9cm, height=3.5cm,
                xlabel={Iteration},
                ylabel={Cost [dB]},
                xlabel style={yshift=1mm},
                ylabel style={yshift=-3mm},
                xtick={0, 100, ..., 500},
                xmin=0, xmax=500,
                ymin=-16, ymax=2,
                ytick={-100, -95, ..., 100},
                grid=both,
                legend pos=north east,
                legend cell align={left},
                tick label style={font=\footnotesize},
                label style={font=\small},
                legend style={font=\small}
            ]
            \addplot[red, opacity=0.4] table [x=Iteration, y = Best_1_Cost_dB] {\data};
            \pgfplotsinvokeforeach{2, 4, 8, 16, 32, 64, 128, 256, 512, 1024}{
                \addplot[red, opacity=0.4, forget plot] table [x=Iteration, y=Best_#1_Cost_dB] {\data};
            }
            \addplot[line width=1.5pt, blue] table [x=Iteration, y=Proxy_Cost_dB] {\data};
            \end{axis}
        \end{tikzpicture}
    }
    \subfloat[$(8\times 1)\times(8\times 1)$]{
        \begin{tikzpicture}
            \pgfplotstableread[col sep=comma]{data/trajectories_8x1_8x1_cost.csv}{\data}
            \begin{axis}[
                width=5.9cm, height=3.5cm,
                xlabel={Iteration},
                ylabel={Cost [dB]},
                xlabel style={yshift=1mm},
                ylabel style={yshift=-3mm},
                xtick={0, 100, ..., 500},
                xmin=0, xmax=500,
                ymin=-13, ymax=3,
                ytick={-100, -95, ..., 100},
                grid=both,
                legend pos=north east,
                legend cell align={left},
                tick label style={font=\footnotesize},
                label style={font=\small},
                legend style={font=\small}
            ]
            \addplot[red, opacity=0.4] table [x=Iteration, y = Best_1_Cost_dB] {\data};
            \pgfplotsinvokeforeach{2, 4, 8, 16, 32, 64, 128, 256, 512, 1024}{
                \addplot[red, opacity=0.4, forget plot] table [x=Iteration, y=Best_#1_Cost_dB] {\data};
            }
            \addplot[line width=1.5pt, blue] table [x=Iteration, y=Proxy_Cost_dB] {\data};
            \end{axis}
        \end{tikzpicture}
    }
    \subfloat[$(32\times 1)\times(32\times 1)$]{
        \begin{tikzpicture}
            \pgfplotstableread[col sep=comma]{data/trajectories_32x1_32x1_cost.csv}{\data}
            \begin{axis}[
                width=5.9cm, height=3.5cm,
                xlabel={Iteration},
                ylabel={Cost [dB]},
                xlabel style={yshift=1mm},
                ylabel style={yshift=-3mm},
                xtick={0, 100, ..., 500},
                xmin=0, xmax=500,
                ymin=-13, ymax=3,
                ytick={-100, -95, ..., 100},
                grid=both,
                legend pos=north east,
                legend cell align={left},
                tick label style={font=\footnotesize},
                label style={font=\small},
                legend style={font=\small}
            ]
            \addplot[red, opacity=0.4] table [x=Iteration, y = Best_1_Cost_dB] {\data};
            \pgfplotsinvokeforeach{2, 4, 8, 16, 32, 64, 128, 256, 512, 1024}{
                \addplot[red, opacity=0.4, forget plot] table [x=Iteration, y=Best_#1_Cost_dB] {\data};
            }
            \addplot[line width=1.5pt, blue] table [x=Iteration, y=Proxy_Cost_dB] {\data};
            \end{axis}
        \end{tikzpicture}
    }\\
    \subfloat[$(2\times 2)\times(2\times 2)$]{
        \begin{tikzpicture}
            \pgfplotstableread[col sep=comma]{data/trajectories_2x2_2x2_cost.csv}{\data}
            \begin{axis}[
                width=5.9cm, height=3.5cm,
                xlabel={Iteration},
                ylabel={Cost [dB]},
                xlabel style={yshift=1mm},
                ylabel style={yshift=-3mm},
                xtick={0, 100, ..., 500},
                xmin=0, xmax=500,
                ymin=-14, ymax=2,
                ytick={-100, -95, ..., 100},
                grid=both,
                legend pos=north east,
                legend cell align={left},
                tick label style={font=\footnotesize},
                label style={font=\small},
                legend style={font=\small}
            ]
            \addplot[red, opacity=0.4] table [x=Iteration, y = Best_1_Cost_dB] {\data};
            \pgfplotsinvokeforeach{2, 4, 8, 16, 32, 64, 128, 256, 512, 1024}{
                \addplot[red, opacity=0.4, forget plot] table [x=Iteration, y=Best_#1_Cost_dB] {\data};
            }
            \addplot[line width=1.5pt, blue] table [x=Iteration, y=Proxy_Cost_dB] {\data};
            \end{axis}
        \end{tikzpicture}
    }
    \subfloat[$(8\times 8)\times(8\times 8)$]{
        \begin{tikzpicture}
            \pgfplotstableread[col sep=comma]{data/trajectories_8x8_8x8_cost.csv}{\data}
            \begin{axis}[
                width=5.9cm, height=3.5cm,
                xlabel={Iteration},
                ylabel={Cost [dB]},
                xlabel style={yshift=1mm},
                ylabel style={yshift=-3mm},
                xtick={0, 100, ..., 500},
                xmin=0, xmax=500,
                ymin=-13, ymax=3,
                ytick={-100, -95, ..., 100},
                grid=both,
                legend pos=north east,
                legend cell align={left},
                tick label style={font=\footnotesize},
                label style={font=\small},
                legend style={font=\small}
            ]
            \addplot[red, opacity=0.4] table [x=Iteration, y = Best_1_Cost_dB] {\data};
            \pgfplotsinvokeforeach{2, 4, 8, 16, 32, 64, 128, 256, 512, 1024}{
                \addplot[red, opacity=0.4, forget plot] table [x=Iteration, y=Best_#1_Cost_dB] {\data};
            }
            \addplot[line width=1.5pt, blue] table [x=Iteration, y=Proxy_Cost_dB] {\data};
            \end{axis}
        \end{tikzpicture}
    }
    \subfloat[$(32\times 32)\times(32\times 32)$]{
        \begin{tikzpicture}
            \pgfplotstableread[col sep=comma]{data/trajectories_32x32_32x32_cost.csv}{\data}
            \begin{axis}[
                width=5.9cm, height=3.5cm,
                xlabel={Iteration},
                ylabel={Cost [dB]},
                xlabel style={yshift=1mm},
                ylabel style={yshift=-3mm},
                xtick={0, 100, ..., 500},
                xmin=0, xmax=500,
                ymin=-13, ymax=3,
                ytick={-100, -95, ..., 100},
                grid=both,
                legend pos=north east,
                legend cell align={left},
                tick label style={font=\footnotesize},
                label style={font=\small},
                legend style={font=\small}
            ]
            \addplot[red, opacity=0.4] table [x=Iteration, y = Best_1_Cost_dB] {\data};
            \pgfplotsinvokeforeach{2, 4, 8, 16, 32, 64, 128, 256, 512, 1024}{
                \addplot[red, opacity=0.4, forget plot] table [x=Iteration, y=Best_#1_Cost_dB] {\data};
            }
            \addplot[line width=1.5pt, blue] table [x=Iteration, y=Proxy_Cost_dB] {\data};
            \end{axis}
        \end{tikzpicture}
    }
    \caption{Gradient descent trajectories (labeled ``random'') with $2^{10}$ random initializations. Among the $2^{10}$ curves, for clarity, only the $\{1,2,2^2,\ldots,2^{10}\}$th best curves are depicted. Also shown (labeled ``genie'') is the proxy of the MLE obtained by initialization at the true geometric parameters, $\br = (0 \text{ m},0 \text{ m}, 10 \text{ m})$ and $\bR = \bI$. Each subfigure's caption describes the setup, i.e., $(\Ntx\times\Nty)\times(\Nrx\times\Nry)$.}
    \label{fig:brute_force_mle}
\end{figure*}
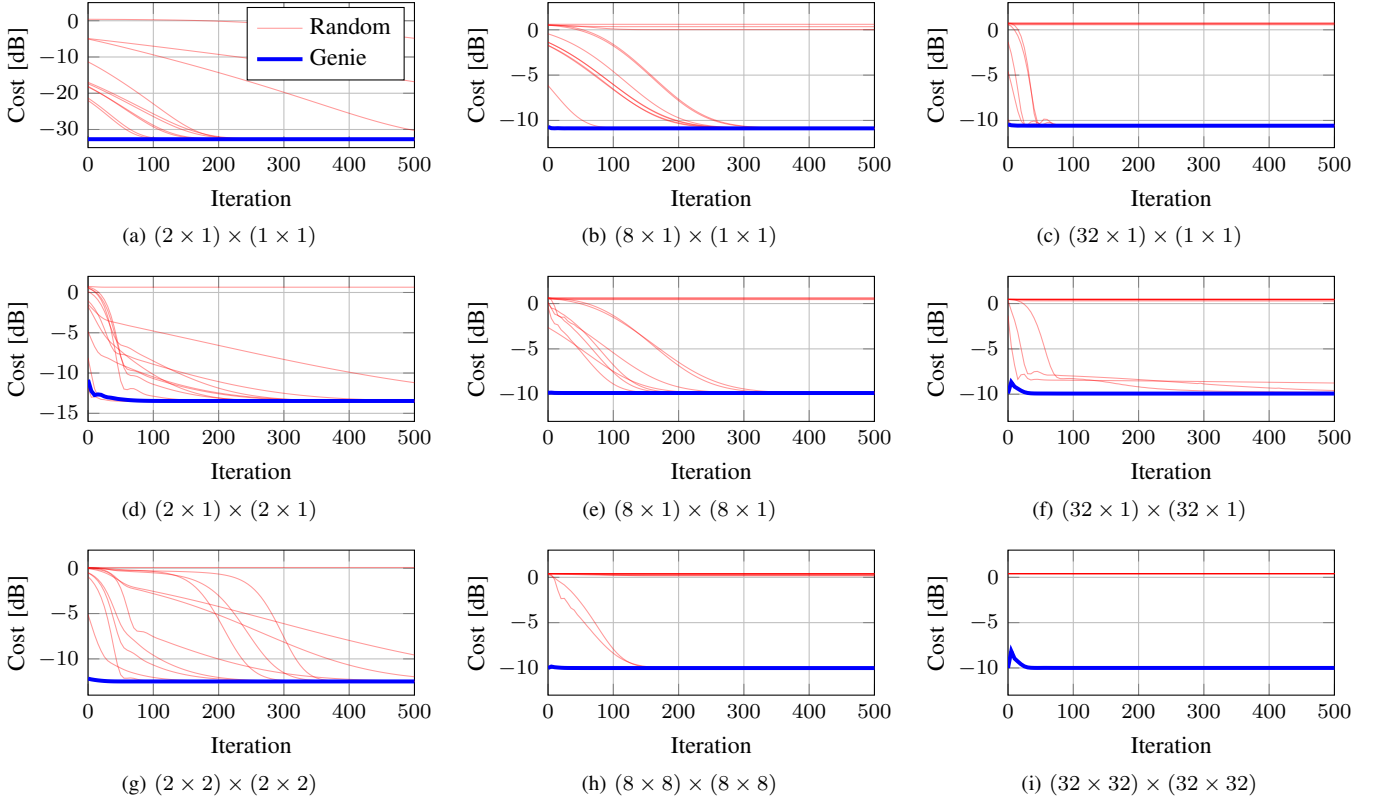

In Fig. \ref{fig:brute_force_mle}, the MLE of the geometric parameters---or, more precisely, its numerical approximation---is tested. Three setups are entertained, involving different types of transmit and receive arrays, always with half-wavelength antenna spacings:
\begin{enumerate}
    \item ULA to single antenna, which entails two geometric parameters.
    \item ULA to ULA, entailing four geometric parameters.
    \item UPA to UPA, entailing six geometric parameters.
\end{enumerate}
The cost function and the MLE implementation are detailed in App. \ref{app:implementation_detail}.
The carrier frequency is 30 GHz, meaning $\lambda_{\rm c}=1$ cm.
For each setup, a channel and noise realizations are generated for $\br = (0 \text{ m},0 \text{ m}, 10 \text{ m})$, $\bR = \bI$ and $\SNR = 10$~dB.
To account for the nonconvexity of the cost function, $2^{10}$ distinct initializations are applied to the gradient-descent algorithm.
As the number of 
parameters and the array dimensions grow large, most initializations fail to converge to the global minimum---the ``genie'' curve serves as its proxy---or, in fact, to converge at all. 

\section{Wavefront Channel Parameterization}
\label{sec:wavefront_parameterization}
 
Letting $\tD_{\nr,\nt}$ be the series expansion of $D_{\nr,\nt}$ with respect to the coordinates,
the quantities to estimate become, for $\bn \in [\bN]$,
\begin{align}
   \!\! \th(\bn) = \exp\! \bigg( \! -j\frac{2\pi }{\lambdac} \tD_{\nr,\nt}\bigg(1 + \df \bigg(\nf- \frac{\Nf-1}{2}\bigg)\!\bigg)\!\bigg) \label{channel_approximate}
\end{align}
in lieu of \eqref{channel}, with $D_{\nr,\nt}\approx D$ and $D_{\nr,\nt}\approx \tD_{\nr,\nt}$ applied to the amplitude and phase, respectively. 
It will be seen next that 
the phase of \eqref{channel_approximate} is polynomial in the indices
$\nrx$, $\nry$, $\ntx$, $\nty$, and $\nf$.

\subsection{Near-Field Channel as a Polynomial Phase Function}

Given the coordinate system set forth in the previous section, 
\begin{align}
    D_{\nr,\nt}
    & = \| \br_\nr-\bt_\nt \| \\
    &= \|\br+((\br_\nr-\br)-\bt_\nt)\|\\
    &=D \, \|\vec{\br}+ \bdelta_{\nr,\nt}\|\\
    &=D \, g(\bdelta_{\nr,\nt}), \label{distance}
\end{align}
where $\vec{\br} \equiv \frac{\br}{D}$ while
\begin{align}
   \bdelta & 
    \equiv \frac{(\br_\nr-\br)-\bt_\nt}{D}\\
   & = \frac{\bR}{D}\!\!
    \begin{bmatrix}
        \drx(\nrx-\frac{\Nrx-1}{2}) \\ \dry(\nry-\frac{\Nry-1}{2}) \\ 0
    \end{bmatrix}\!\!-\!
    \frac{1}{D}\!\!
    \begin{bmatrix}
        \dtx(\ntx-\frac{\Ntx-1}{2}) \\ \dty(\nty-\frac{\Nty-1}{2}) \\ 0
    \end{bmatrix} \label{delta_definition}
\end{align}
and $g(\bdelta) \equiv \|\vec{\br}+ \bdelta\|$.
Using
\begin{align}
    \sqrt{1+2xt + t^2}
    & = \sum_\ell {{\scriptstyle 1 / 2} \choose \ell} (2xt + t^2)^\ell\\
    & = 1+ (2xt + t^2) - \frac{1}{8}(2xt + t^2)^2\\
    &\quad + \frac{1}{16}(2xt + t^2)^3- \frac{5}{128}(2xt + t^2)^4 + \cdots \nonumber\\
    &= 1+xt + \frac{1}{2}(1-x^2)t^2 - \frac{1}{2}(x-x^3)t^3 \nonumber\\
    &\quad  - \frac{1}{8}(1-6x^2+5x^4)t^4 + \cdots, 
\end{align}
valid for
\begin{align}
t<\min \! \left(|-x-\sqrt{x^2-1}|,|-x+\sqrt{x^2-1}|\right),    
\end{align}
one obtains \eqref{power_series}; see App. \ref{app:legendre} for a closed-form expression involving Legendre polynomials.
\begin{figure*}
\begin{align}
    g(\bdelta)
    & = \sqrt{1+ 2\vec{\br}^{\vphantom{\big(}\top}\vec{\bdelta}\cdot\|\bdelta\| + \|\bdelta\|^2}\\
    & = 1 + \vec{\br}^\top\vec{\bdelta}\cdot\|\bdelta\| + \frac{1}{2}\big(1-(\vec{\br}^\top\vec{\bdelta})^2\big)\|\bdelta\|^2 - \frac{1}{2}\big(\vec{\br}^\top\vec{\bdelta}-(\vec{\br}^\top\vec{\bdelta})^3\big) \|\bdelta\|^3 - \frac{1}{8}\big(1- 6(\vec{\br}^\top\vec{\bdelta})^2 + 5(\vec{\br}^\top\vec{\bdelta})^4 \big)\|\bdelta\|^4 + \cdots\\
    & = \underbrace{ \underbrace{ \underbrace{ 1 + \vec{\br}^\top\bdelta }_{=g_1(\bdelta)} + \frac{1}{2}\big(\|\bdelta\|^2-(\vec{\br}^\top\bdelta)^2\big) }_{=g_2(\bdelta)} - \frac{1}{2}\big(\vec{\br}^\top\bdelta\cdot \|\bdelta\|^2-(\vec{\br}^\top\bdelta)^3\big) - \frac{1}{8}\big(\|\bdelta\|^4- 6(\vec{\br}^\top\bdelta)^2\|\bdelta\|^2 + 5(\vec{\br}^\top\bdelta)^4 \big)}_{=g_4(\bdelta)} + \cdots \label{power_series}
\end{align}
\hrulefill
\end{figure*}
Discarding terms with degree higher than $L$ yields a Taylor polynomial of degree $L$, denoted by $g_L(\bdelta)$.
Particularly relevant are the cases $L=1$ and $L=2$, which respectively correspond to planar and parabolic wavefront models \cite{do2023parabolic}.


\begin{table}
\setlength{\tabcolsep}{3pt}
\addtolength{\leftskip} {-2cm}
\addtolength{\rightskip} {-2cm}
\centering
\begin{threeparttable}
\caption{Comparison of the Geometric and \\ Wavefront Parameterizations}
\label{table:number_of_parameters_comparison}
\begin{tabular}{c c | c | c c c c} 
\toprule
 &  & \# geometric & \multicolumn{4}{c}{\# polynomial coefficients}
\\
Tx array & Rx array & parameters & General & $L=1$ & $L=2$ & $L=3$
\\
\midrule
linear & single & 2 & ${L+1 \choose 1}$ & 2 & 3 & 4 \\[2pt]
planar & single & 3 & ${L+2 \choose 2}$ & 3 & 6 & 10 \\[2pt]
linear & linear & 4 & ${L+2 \choose 2}$ & 3 & 6 & 10\\[2pt]
planar & linear & 5 & ${L+3 \choose 3}$ & 4 & 10 & 20\\[2pt]
planar & planar & 6 & ${L+4 \choose 4}$ & 5 & 15 & 35\\[2pt]
\bottomrule
\end{tabular}
\end{threeparttable}
\end{table}

From \eqref{delta_definition}, each component of $\bdelta$ is an affine function of  $\nrx$, $\nry$, $\ntx$, and $\nty$. 
Thus, $\tD_{\nr,\nt} = D \, g_L(\bdelta_{\nr,\nt})$ is a multidimensional (multivariate) polynomial of degree $L$ in these indices.
The set of degrees, $\cL$, 
is determined by $L$ and by the array dimensionalities. For transmit and receive UPAs,
\begin{align}
    \cL = \{(\ell_1,\ell_2,\ell_3,\ell_4) \in\bbN_0^4: \ell_1+\ell_2+\ell_3+\ell_4 \leq L\} \label{degree_set}
\end{align}
whose cardinality is $|\cL| = {L+4 \choose 4}$ \cite{chen1992principles}.
If one of the UPAs degenerates into a ULA,
the degree for the singleton dimension becomes zero;
for instance, if $\Nty=1$, then
\begin{align}
    \cL & = \{\overbrace{(0,0,0)}^{\text{degree }0}, \overbrace{(0,0,1), (0,1,0), (1,0,0)}^{\text{degree }1} , \nonumber \\
    & \qquad (0,0,2),(0,2,0),(2,0,0),(0,1,1),(1,0,1),(1,1,0), \nonumber \\
    &\qquad \ldots, \underbrace{(0,0,L), \ldots, (L,0,0)}_{\text{degree }L}\} \times \{0\}
\end{align}
and $|\cL| = {L+3 \choose 3}$.
Finally, the phase of the wavefront parametric model in
\eqref{channel_approximate} is a multidimensional polynomial of $\nrx$, $\nry$, $\ntx$, $\nty$, and $\nf$, with a set of degrees $\cM \equiv \cL\times\{0,1\}$.

A comparison of the parameter counts in the geometric and wavefront parameterizations is provided in Table \ref{table:number_of_parameters_comparison}.

\subsection{Model Mismatch}

The modelling error in the polynomial expansion of the spherical wavefront
is dominated by the lowest-degree truncated term.
The ensuing MSE floor is $\cO(\|\bdelta\|^{2(L+1)})$, quickly decaying as $L$ grows,
with $L=2$ or $L=3$ sufficing for a high fidelity over a very wide range of conditions.

For $L=1$ specifically, the dominant error term can be bounded by \cite[Ch. 4.4]{balanis2016antenna}
\begin{align}
    \frac{2\pi D}{\lambda} \cdot \frac{1}{2}\big|1-(\vec{\br}^\top\vec{\bdelta})^2\big|\|\bdelta\|^2
    &\leq \pi \frac{D}{\lambda} \|\bdelta\|^2 \label{Venice1}
\end{align}
with equality for 
$\vec{\br}^\top\bdelta = 0$.
In turn, for $L=2$,
\begin{align}
    \frac{2\pi D}{\lambda} \cdot \frac{1}{2}\big|\vec{\br}^\top\vec{\bdelta}-(\vec{\br}^\top\vec{\bdelta})^3\big| \|\bdelta\|^3
    &\leq \frac{2\pi}{3\sqrt{3}} \frac{D}{\lambda} \|\bdelta\|^3 \label{Venice2}
\end{align}
with equality for
$\vec{\br}^\top\vec{\bdelta} = \pm \frac{1}{\sqrt{3}}\|\bdelta\|$.
And, for $L=3$,
\begin{align}
    &\frac{2\pi D}{\lambda} \cdot\frac{1}{8}\big(1- 6(\vec{\br}^\top\vec{\bdelta})^2 + 5(\vec{\br}^\top\vec{\bdelta})^4 \big)\|\bdelta\|^4
 \leq \frac{\pi}{4} \frac{D}{\lambda} \|\bdelta\|^4 \label{Venice3}
\end{align}
with equality for
$\vec{\br}^\top\bdelta = 0$.

The worst-case error, corresponding to the worst orientation and worst antenna pair, can be obtained by means of the triangle inequality as
\begin{align}
    \|\bdelta \| &= \frac{\|(\br_\nr-\br)-\bt_\nt\|}{D}\\
    &\leq \frac{\|\br_\nr-\br\|+\|\bt_\nt\|}{D}\\
    &\leq \frac{L_{\rm t}+L_{\rm r}}{2D}, \label{PSG}
\end{align}
where $L_{\rm t}$ and $L_{\rm r}$ are the largest dimensions of the transmit and receive arrays, respectively (the diagonal lengths in the case of UPAs). Combining \eqref{PSG} with \eqref{Venice1}, \eqref{Venice2}, and \eqref{Venice3}, the worst-case errors for $L=1$, $2$, and $3$ emerge as
\begin{align}
    \frac{\pi}{4} \frac{(L_{\rm t}+L_{\rm r})^2}{\lambda D} \quad\,\,\, \frac{\pi}{12\sqrt{3}} \frac{(L_{\rm t}+L_{\rm r})^3}{\lambda D^2} \quad\,\,\, \frac{\pi}{64} \frac{(L_{\rm t}+L_{\rm r})^4}{\lambda D^3}.
\end{align}
Setting the worst-case error to $\pi/8$ for $L=1$ returns the Fraunhofer distance expression
\begin{align}
    D = \frac{2 \, (L_{\rm t}+L_{\rm r})^2}{\lambda}, \label{fraunhofer}
\end{align}
which is often taken to delineate the boundary between the near- and far-field regions.
The $\pi/8$ threshold, however, is somewhat arbitrary and the necessary accuracy depends on the task at hand \cite{sun2025how}. For channel estimation in particular, the mismatch measured in per-entry MSE is roughly the squared chord length, $(2\sin\frac{\pi}{16})^2 \approx -8.17$ dB, and a higher accuracy is often needed (refer to Sec. \ref{sec:numerical_evaluation} for a follow-up discussion on the matter).

\section{Interlude: Multidimensional Polynomial Phase Estimation}
\label{sec:polynomial_phase_estimation}

With a suitable choice of $L$, the observations in \eqref{observation_model} can be approximated to any desired accuracy, for $\bn\in[\bN]$,
by
\begin{align}
   \tilde{h}(\bn) + w_{\bbC}(\bn) \label{approximateObservationModel}
\end{align}
with $\tilde{h}(\bn)$ given in \eqref{channel_approximate}.
With the near-field LOS channel estimation thus cast as a multidimensional polynomial phase estimation problem, the method developed in \cite{do2025multidimensional} can be applied. For the sake of completeness, a summary of this method is provided in this section.

\subsection{Polynomial Phase Model}

Consider the $\sfD$-dimensional function $e^{j2\pi x(\bn)}$, defined over $\bn=(n_0,\ldots,n_{\sfD-1})$, where
\begin{align}
    x(\bn)&= \sum_{\bmm} 
 a_{\bmm}p_\bmm(\bn)  \label{polynomialMonomial}
\end{align}
with $\bmm=(m_0,\ldots,m_{\sfD-1})$ and 
\begin{align}
    p_\bmm(\bn) = \frac{\bn^\bmm}{\bmm!} + (\text{lower-degree terms}).
\end{align}
The coefficient $a_\bmm$ vanishes outside the set of polynomial degrees, $\cM$.
The noisy observation is
\begin{align}
    y(\bn) = e^{j2\pi x(\bn)}  + w_{\bbC}(\bn), \label{signalModel}
\end{align}
of which \eqref{approximateObservationModel} is an instance. 
The goal is to estimate the polynomial coefficients $\{ a_\bmm \}$ from $y(\bn)$.

To ensure that the set of polynomials $\{p_\bmm(\bn)\}$ is not overloaded, it is necessary that \cite[Sec. II-B]{do2025multidimensional}
\begin{equation}
    \bmm\subset [\bN] .
    \label{Ormuz}
\end{equation}
For the problem at hand, this maps to $L<N_{x}$ and $L<N_{y}$, where $N_x$ and $N_y$ denote the number of observations on the two dimensions of the transmit array. Thus, if the transmit array is linear, at least $(L+1)$ pilot observations are required; if the array is planar, it is $(L+1)^2$.


\begin{algorithm}[t]
\caption{Multidimensional Polynomial Phase Estimation}\label{alg:proposed}
\begin{algorithmic}
\Procedure{Estimate-Coefficients}{$y, \cM$}
    \For{$\bmm\in\cM$ (in descending order)}
    \State $\hat{a}_{\bmm} \gets \frac{1}{2\pi}\arg(\mu_{\bmm}(\cD^{\bmm} y))$
    \State $y(\bn) \gets y(\bn)\exp\!\big(\!-j2\pi \hat{a}_{\bmm}p_\bmm(\bn) \big)$
    \EndFor
    \State \Return $\{\hat{a}_\bmm\}$
\EndProcedure
\end{algorithmic}
\end{algorithm}

\subsection{Estimation Method}

Summarized in Alg. \ref{alg:proposed}, the method in \cite{do2025multidimensional}
attains the CRB at high SNR.
In it,
$\cD^{\bmm} \equiv \cD_0^{m_0} \cdots \cD_{\sfD-1}^{m_{\sfD-1}}$
where
\begin{align}
(\cD_d s)(\bn) \equiv s(\bn+\bee_d)\overline{s(\bn)}    
\end{align}
with $\bee_d$ the $d$th standard unit vector. In turn,
\begin{align}
    \mu_\bmm(s) & = \Pi\bigg[\sum_\bell (\Pi s)(\bell)\bigg] \label{proposedAverager}\\
    &\quad \cdot \exp\!\bigg(j  \sum_\bn u_{\bmm}(\bn)  \arg\!\bigg(s(\bn)\overline{\sum_\bell (\Pi s)(\bell)}\bigg)\bigg) , \nonumber 
\end{align}
where $(\Pi s)(\bn) = e^{j\arg(s(\bn))}$ for $s(\bn) \neq 0$,
while
\begin{align}
    u_{\bmm}(\bn) = \frac{{\bn+\bmm \choose \bmm}{\bN-\bn-\one \choose \bmm}}{{\bN+\bmm\choose 2\bmm+\one}}  \qquad\quad \bn\in[\bN-\bmm] .
    \label{kayWeight}
\end{align}
The complexity is strictly linear in $|[\bN]|$ (which corresponds to linear in $N_{{\rm t},x}N_{{\rm t},y} N_{{\rm r},x} N_{{\rm r},y} \Nf$ for the problem at hand),
and subquadratic in the number of polynomial coefficients. 

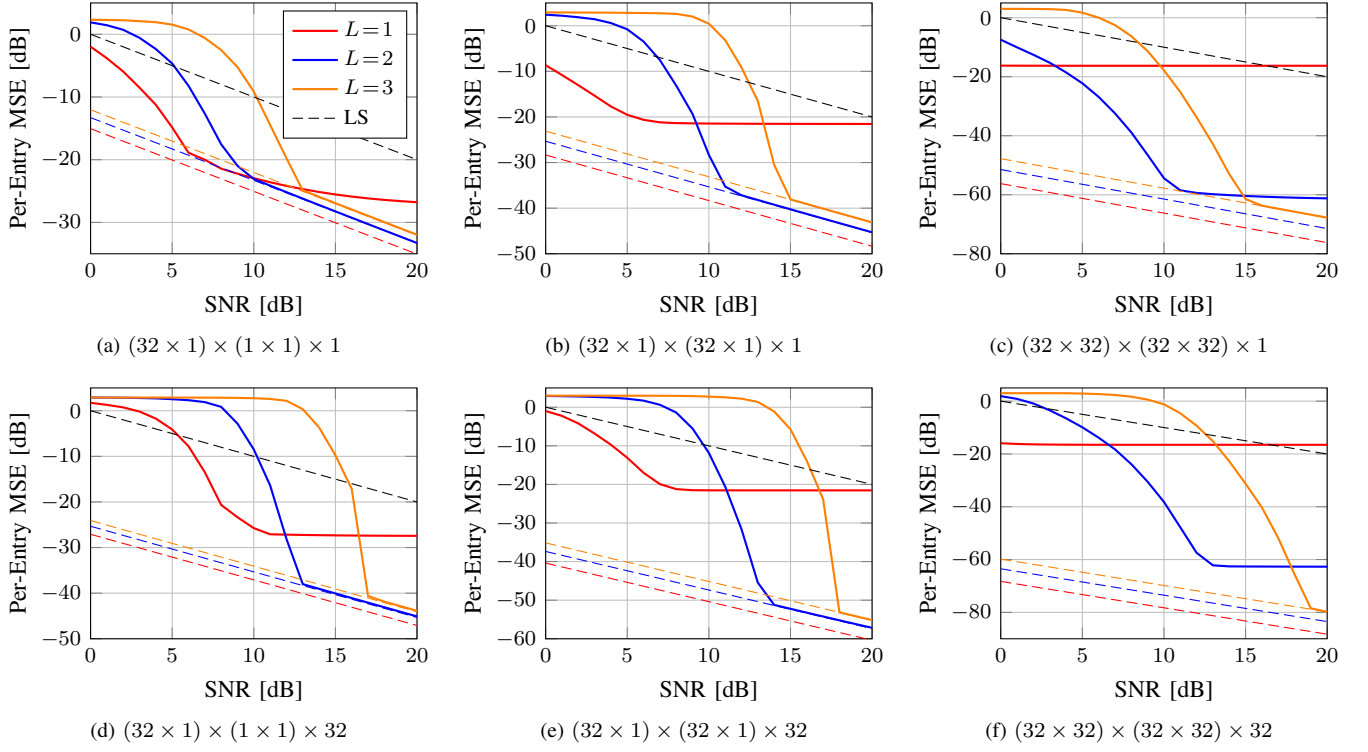
\begin{figure*}
    \centering
        \subfloat[$(32\times 1)\times(1\times 1)\times 1$]{
        \begin{tikzpicture}
            \pgfplotstableread[col sep=comma]{data/simulation_results_32x1_1x1_Nf1_unit_amp.csv}{\data}
            \begin{axis}[
                width=5.9cm, height=4.9cm,
                xlabel={SNR [dB]},
                ylabel={Per-Entry MSE [dB]},
                xlabel style={yshift=1mm},
                ylabel style={yshift=-3mm},
                xtick={0, 5, ..., 20},
                xmin=0, xmax=20,
                ymin=-35, ymax=5,
                ytick={-40, -30, ..., 0},
                grid=both,
                legend pos=north east,
                legend cell align={left},
                tick label style={font=\footnotesize},
                label style={font=\small},
                legend style={font=\footnotesize}
            ]
            \addplot[red, thick] table [x=snr_db, y = mse_db_1] {\data};
            \addlegendentry{$L\!=\!1$}
            \addplot[blue, thick] table [x=snr_db, y = mse_db_2] {\data};
            \addlegendentry{$L\!=\!2$}
            \addplot[orange, thick] table [x=snr_db, y = mse_db_3] {\data};
            \addlegendentry{$L\!=\!3$}
            
            \addplot[black, dash pattern=on 4pt off 2pt, domain=0:20] {-x};
            \addlegendentry{LS}
            
            \addplot[red, dash pattern=on 4pt off 2pt, domain=0:20] {10*log10(2/(2*10^(x/10))/32)};
            
            \addplot[blue, dash pattern=on 4pt off 2pt, domain=0:20] {10*log10(3/(2*10^(x/10))/32)};
            
            \addplot[orange, dash pattern=on 4pt off 2pt, domain=0:20] {10*log10(4/(2*10^(x/10))/32)};

            \end{axis}
        \end{tikzpicture}
    }
    \subfloat[$(32\times 1)\times(32\times 1)\times 1$]{
        \begin{tikzpicture}
            \pgfplotstableread[col sep=comma]{data/simulation_results_32x1_32x1_Nf1_unit_amp.csv}{\data}
            \begin{axis}[
                width=5.9cm, height=4.9cm,
                xlabel={SNR [dB]},
                ylabel={Per-Entry MSE [dB]},
                xlabel style={yshift=1mm},
                ylabel style={yshift=-3mm},
                xtick={0, 5, ..., 20},
                xmin=0, xmax=20,
                ymin=-50, ymax=5,
                ytick={-50, -40, ..., 0},
                grid=both,
                legend pos=north east,
                legend cell align={left},
                tick label style={font=\footnotesize},
                label style={font=\small},
                legend style={font=\footnotesize}
            ]
            \addplot[red, thick] table [x=snr_db, y = mse_db_1] {\data};
            \addplot[blue, thick] table [x=snr_db, y = mse_db_2] {\data};
            \addplot[orange, thick] table [x=snr_db, y = mse_db_3] {\data};

            \addplot[black, dash pattern=on 4pt off 2pt, domain=0:20] {-x};
            
            \addplot[red, dash pattern=on 4pt off 2pt, domain=0:20] {10*log10(3/(2*10^(x/10))/(32*32))};
            
            \addplot[blue, dash pattern=on 4pt off 2pt, domain=0:20] {10*log10(6/(2*10^(x/10))/(32*32))};
            
            \addplot[orange, dash pattern=on 4pt off 2pt, domain=0:20] {10*log10(10/(2*10^(x/10))/(32*32))};
            
            \end{axis}
        \end{tikzpicture}
    }
    \subfloat[$(32\times 32)\times(32\times 32)\times 1$]{
        \begin{tikzpicture}
            \pgfplotstableread[col sep=comma]{data/simulation_results_32x32_32x32_Nf1_unit_amp.csv}{\data}
            \begin{axis}[
                width=5.9cm, height=4.9cm,
                xlabel={SNR [dB]},
                ylabel={Per-Entry MSE [dB]},
                xlabel style={yshift=1mm},
                ylabel style={yshift=-3mm},
                xtick={0, 5, ..., 20},
                xmin=0, xmax=20,
                ymin=-80, ymax=5,
                ytick={-80, -60, ..., 0},
                grid=both,
                legend pos=north east,
                legend cell align={left},
                tick label style={font=\footnotesize},
                label style={font=\small},
                legend style={font=\small}
            ] 
            \addplot[red, thick] table [x=snr_db, y = mse_db_1] {\data};
            \addplot[blue, thick] table [x=snr_db, y = mse_db_2] {\data};
            \addplot[orange, thick] table [x=snr_db, y = mse_db_3] {\data};

            \addplot[black, dash pattern=on 4pt off 2pt, domain=0:20] {-x};
            
            \addplot[red, dash pattern=on 4pt off 2pt, domain=0:20] {10*log10(5/(2*10^(x/10))/(32*32*32*32))};
            
            \addplot[blue, dash pattern=on 4pt off 2pt, domain=0:20] {10*log10(15/(2*10^(x/10))/(32*32*32*32))};
            
            \addplot[orange, dash pattern=on 4pt off 2pt, domain=0:20] {10*log10(35/(2*10^(x/10))/(32*32*32*32))};
            
            \end{axis}
        \end{tikzpicture}
    }\\
    \subfloat[$(32\times 1)\times(1\times 1)\times 32$]{
        \begin{tikzpicture}
            \pgfplotstableread[col sep=comma]{data/simulation_results_32x1_1x1_Nf32_unit_amp.csv}{\data}
            \begin{axis}[
                width=5.9cm, height=4.9cm,
                xlabel={SNR [dB]},
                ylabel={Per-Entry MSE [dB]},
                xlabel style={yshift=1mm},
                ylabel style={yshift=-3mm},
                xtick={0, 5, ..., 20},
                xmin=0, xmax=20,
                ymin=-50, ymax=5,
                ytick={-80, -70, ..., 0},
                grid=both,
                legend pos=north east,
                legend cell align={left},
                tick label style={font=\footnotesize},
                label style={font=\small},
                legend style={font=\small}
            ]
            \addplot[red, thick] table [x=snr_db, y = mse_db_1] {\data};
            \addplot[blue, thick] table [x=snr_db, y = mse_db_2] {\data};
            \addplot[orange, thick] table [x=snr_db, y = mse_db_3] {\data};

            \addplot[black, dash pattern=on 4pt off 2pt, domain=0:20] {-x};
            
            \addplot[red, dash pattern=on 4pt off 2pt, domain=0:20] {10*log10(2*2/(2*10^(x/10))/(32*32))};
            
            \addplot[blue, dash pattern=on 4pt off 2pt, domain=0:20] {10*log10(2*3/(2*10^(x/10))/(32*32))};
            
            \addplot[orange, dash pattern=on 4pt off 2pt, domain=0:20] {10*log10(2*4/(2*10^(x/10))/(32*32))};
            
            \end{axis}
        \end{tikzpicture}
    }
    \subfloat[$(32\times 1)\times(32\times 1)\times 32$]{
        \begin{tikzpicture}
            \pgfplotstableread[col sep=comma]{data/simulation_results_32x1_32x1_Nf32_unit_amp.csv}{\data}
            \begin{axis}[
                width=5.9cm, height=4.9cm,
                xlabel={SNR [dB]},
                ylabel={Per-Entry MSE [dB]},
                xlabel style={yshift=1mm},
                ylabel style={yshift=-3mm},
                xtick={0, 5, ..., 20},
                xmin=0, xmax=20,
                ymin=-60, ymax=5,
                ytick={-80, -70, ..., 0},
                grid=both,
                legend pos=north east,
                legend cell align={left},
                tick label style={font=\footnotesize},
                label style={font=\small},
                legend style={font=\small}
            ]
            \addplot[red, thick] table [x=snr_db, y = mse_db_1] {\data};
            \addplot[blue, thick] table [x=snr_db, y = mse_db_2] {\data};
            \addplot[orange, thick] table [x=snr_db, y = mse_db_3] {\data};

            \addplot[black, dash pattern=on 4pt off 2pt, domain=0:20] {-x};
            
            \addplot[red, dash pattern=on 4pt off 2pt, domain=0:20] {10*log10(2*3/(2*10^(x/10))/(32*32*32))};
            
            \addplot[blue, dash pattern=on 4pt off 2pt, domain=0:20] {10*log10(2*6/(2*10^(x/10))/(32*32*32))};
            
            \addplot[orange, dash pattern=on 4pt off 2pt, domain=0:20] {10*log10(2*10/(2*10^(x/10))/(32*32*32))};
            
            \end{axis}
        \end{tikzpicture}
    }
    \subfloat[$(32\times 32)\times(32\times 32)\times 32$]{
        \begin{tikzpicture}
            \pgfplotstableread[col sep=comma]{data/simulation_results_32x32_32x32_Nf32_unit_amp.csv}{\data}
            \begin{axis}[
                width=5.9cm, height=4.9cm,
                xlabel={SNR [dB]},
                ylabel={Per-Entry MSE [dB]},
                xlabel style={yshift=1mm},
                ylabel style={yshift=-3mm},
                xtick={0, 5, ..., 20},
                xmin=0, xmax=20,
                ymin=-90, ymax=5,
                ytick={-100, -80, ..., 0},
                grid=both,
                legend pos=north east,
                legend cell align={left},
                tick label style={font=\footnotesize},
                label style={font=\small},
                legend style={font=\small}
            ] 
            \addplot[red, thick] table [x=snr_db, y = mse_db_1] {\data};
            \addplot[blue, thick] table [x=snr_db, y = mse_db_2] {\data};
            \addplot[orange, thick] table [x=snr_db, y = mse_db_3] {\data};

            \addplot[black, dash pattern=on 4pt off 2pt, domain=0:20] {-x};
            
            \addplot[red, dash pattern=on 4pt off 2pt, domain=0:20] {10*log10(2*5/(2*10^(x/10))/(32*32*32*32*32))};
            
            \addplot[blue, dash pattern=on 4pt off 2pt, domain=0:20] {10*log10(2*15/(2*10^(x/10))/(32*32*32*32*32))};
            
            \addplot[orange, dash pattern=on 4pt off 2pt, domain=0:20] {10*log10(2*35/(2*10^(x/10))/(32*32*32*32*32))};
            
            \end{axis}
        \end{tikzpicture}
    }
    \caption{Channel estimation MSE for the various setups. In the channel model, the amplitude term $D/D_{\nr,\nt}$ in \eqref{channel} is set to $1$ to exclude the effect of amplitude variations. Also shown in dashed are the asymptotes in \eqref{per_entry_crb}. 
    The results are averaged over 1000 realizations of the geometric parameters and the noise (100 realizations for the last subfigure). Each subfigure's caption describes the setup as $(\Ntx\times\Nty)\times(\Nrx\times\Nry)\times \Nf$.
    }
    \label{fig:mse_results_unit}
\end{figure*}

\begin{figure*}
    \centering
    \subfloat[$(32\times 1)\times(1\times 1)\times 1$]{
        \begin{tikzpicture}
            \pgfplotstableread[col sep=comma]{data/simulation_results_32x1_1x1_Nf1_actual_amp.csv}{\data}
            \begin{axis}[
                width=5.9cm, height=4.9cm,
                xlabel={SNR [dB]},
                ylabel={Per-Entry MSE [dB]},
                xlabel style={yshift=1mm},
                ylabel style={yshift=-3mm},
                xtick={0, 5, ..., 20},
                xmin=0, xmax=20,
                ymin=-35, ymax=5,
                ytick={-40, -30, ..., 0},
                grid=both,
                legend pos=north east,
                legend cell align={left},
                tick label style={font=\footnotesize},
                label style={font=\small},
                legend style={font=\footnotesize}
            ]
            \addplot[red, thick] table [x=snr_db, y = mse_db_1] {\data};
            \addlegendentry{$L\!=\!1$}
            \addplot[blue, thick] table [x=snr_db, y = mse_db_2] {\data};
            \addlegendentry{$L\!=\!2$}
            \addplot[orange, thick] table [x=snr_db, y = mse_db_3] {\data};
            \addlegendentry{$L\!=\!3$}
            
            \addplot[black, dash pattern=on 4pt off 2pt, domain=0:20] {-x};
            \addlegendentry{LS}
            
            \addplot[red, dash pattern=on 4pt off 2pt, domain=0:20] {10*log10(2/(2*10^(x/10))/32)};
            
            \addplot[blue, dash pattern=on 4pt off 2pt, domain=0:20] {10*log10(3/(2*10^(x/10))/32)};
            
            \addplot[orange, dash pattern=on 4pt off 2pt, domain=0:20] {10*log10(4/(2*10^(x/10))/32)};

            \end{axis}
        \end{tikzpicture}
    }
    \subfloat[$(32\times 1)\times(32\times 1)\times 1$]{
        \begin{tikzpicture}
            \pgfplotstableread[col sep=comma]{data/simulation_results_32x1_32x1_Nf1_actual_amp.csv}{\data}
            \begin{axis}[
                width=5.9cm, height=4.9cm,
                xlabel={SNR [dB]},
                ylabel={Per-Entry MSE [dB]},
                xlabel style={yshift=1mm},
                ylabel style={yshift=-3mm},
                xtick={0, 5, ..., 20},
                xmin=0, xmax=20,
                ymin=-50, ymax=5,
                ytick={-50, -40, ..., 0},
                grid=both,
                legend pos=north east,
                legend cell align={left},
                tick label style={font=\footnotesize},
                label style={font=\small},
                legend style={font=\small}
            ]
            \addplot[red, thick] table [x=snr_db, y = mse_db_1] {\data};
            \addplot[blue, thick] table [x=snr_db, y = mse_db_2] {\data};
            \addplot[orange, thick] table [x=snr_db, y = mse_db_3] {\data};

            \addplot[black, dash pattern=on 4pt off 2pt, domain=0:20] {-x};
            
            \addplot[red, dash pattern=on 4pt off 2pt, domain=0:20] {10*log10(3/(2*10^(x/10))/(32*32))};
            
            \addplot[blue, dash pattern=on 4pt off 2pt, domain=0:20] {10*log10(6/(2*10^(x/10))/(32*32))};
            
            \addplot[orange, dash pattern=on 4pt off 2pt, domain=0:20] {10*log10(10/(2*10^(x/10))/(32*32))};
            
            \end{axis}
        \end{tikzpicture}
    }
    \subfloat[$(32\times 32)\times(32\times 32)\times 1$]{
        \begin{tikzpicture}
            \pgfplotstableread[col sep=comma]{data/simulation_results_32x32_32x32_Nf32_actual_amp.csv}{\data}
            \begin{axis}[
                width=5.9cm, height=4.9cm,
                xlabel={SNR [dB]},
                ylabel={Per-Entry MSE [dB]},
                xlabel style={yshift=1mm},
                ylabel style={yshift=-3mm},
                xtick={0, 5, ..., 20},
                xmin=0, xmax=20,
                ymin=-80, ymax=5,
                ytick={-80, -60, ..., 0},
                grid=both,
                legend pos=north east,
                legend cell align={left},
                tick label style={font=\footnotesize},
                label style={font=\small},
                legend style={font=\small}
            ]
            \addplot[red, thick] table [x=snr_db, y = mse_db_1] {\data};
            \addplot[blue, thick] table [x=snr_db, y = mse_db_2] {\data};
            \addplot[orange, thick] table [x=snr_db, y = mse_db_3] {\data};

            \addplot[black, dash pattern=on 4pt off 2pt, domain=0:20] {-x};
            
            \addplot[red, dash pattern=on 4pt off 2pt, domain=0:20] {10*log10(2*5/(2*10^(x/10))/(32*32*32*32))};
            
            \addplot[blue, dash pattern=on 4pt off 2pt, domain=0:20] {10*log10(2*15/(2*10^(x/10))/(32*32*32*32))};
            
            \addplot[orange, dash pattern=on 4pt off 2pt, domain=0:20] {10*log10(2*35/(2*10^(x/10))/(32*32*32*32))};
            \end{axis}
        \end{tikzpicture}
    }\\
    \subfloat[$(32\times 1)\times(1\times 1)\times 32$]{
        \begin{tikzpicture}
            \pgfplotstableread[col sep=comma]{data/simulation_results_32x1_1x1_Nf32_actual_amp.csv}{\data}
            \begin{axis}[
                width=5.9cm, height=4.9cm,
                xlabel={SNR [dB]},
                ylabel={Per-Entry MSE [dB]},
                xlabel style={yshift=1mm},
                ylabel style={yshift=-3mm},
                xtick={0, 5, ..., 20},
                xmin=0, xmax=20,
                ymin=-50, ymax=5,
                ytick={-80, -70, ..., 0},
                grid=both,
                legend pos=north east,
                legend cell align={left},
                tick label style={font=\footnotesize},
                label style={font=\small},
                legend style={font=\small}
            ]
            \addplot[red, thick] table [x=snr_db, y = mse_db_1] {\data};
            \addplot[blue, thick] table [x=snr_db, y = mse_db_2] {\data};
            \addplot[orange, thick] table [x=snr_db, y = mse_db_3] {\data};

            \addplot[black, dash pattern=on 4pt off 2pt, domain=0:20] {-x};
            
            \addplot[red, dash pattern=on 4pt off 2pt, domain=0:20] {10*log10(2*2/(2*10^(x/10))/(32*32))};
            
            \addplot[blue, dash pattern=on 4pt off 2pt, domain=0:20] {10*log10(2*3/(2*10^(x/10))/(32*32))};
            
            \addplot[orange, dash pattern=on 4pt off 2pt, domain=0:20] {10*log10(2*4/(2*10^(x/10))/(32*32))};
            
            \end{axis}
        \end{tikzpicture}
    }
    \subfloat[$(32\times 1)\times(32\times 1)\times 32$]{
        \begin{tikzpicture}
            \pgfplotstableread[col sep=comma]{data/simulation_results_32x1_32x1_Nf32_actual_amp.csv}{\data}
            \begin{axis}[
                width=5.9cm, height=4.9cm,
                xlabel={SNR [dB]},
                ylabel={Per-Entry MSE [dB]},
                xlabel style={yshift=1mm},
                ylabel style={yshift=-3mm},
                xtick={0, 5, ..., 20},
                xmin=0, xmax=20,
                ymin=-60, ymax=5,
                ytick={-80, -70, ..., 0},
                grid=both,
                legend pos=north east,
                legend cell align={left},
                tick label style={font=\footnotesize},
                label style={font=\small},
                legend style={font=\small}
            ]
            \addplot[red, thick] table [x=snr_db, y = mse_db_1] {\data};
            \addplot[blue, thick] table [x=snr_db, y = mse_db_2] {\data};
            \addplot[orange, thick] table [x=snr_db, y = mse_db_3] {\data};

            \addplot[black, dash pattern=on 4pt off 2pt, domain=0:20] {-x};
            
            \addplot[red, dash pattern=on 4pt off 2pt, domain=0:20] {10*log10(2*3/(2*10^(x/10))/(32*32*32))};
            
            \addplot[blue, dash pattern=on 4pt off 2pt, domain=0:20] {10*log10(2*6/(2*10^(x/10))/(32*32*32))};
            
            \addplot[orange, dash pattern=on 4pt off 2pt, domain=0:20] {10*log10(2*10/(2*10^(x/10))/(32*32*32))};
            
            \end{axis}
        \end{tikzpicture}
    }
    \subfloat[$(32\times 32)\times(32\times 32)\times 32$]{
        \begin{tikzpicture}
            \pgfplotstableread[col sep=comma]{data/simulation_results_32x32_32x32_Nf1_actual_amp.csv}{\data}
            \begin{axis}[
                width=5.9cm, height=4.9cm,
                xlabel={SNR [dB]},
                ylabel={Per-Entry MSE [dB]},
                xlabel style={yshift=1mm},
                ylabel style={yshift=-3mm},
                xtick={0, 5, ..., 20},
                xmin=0, xmax=20,
                ymin=-90, ymax=5,
                ytick={-100, -80, ..., 0},
                grid=both,
                legend pos=north east,
                legend cell align={left},
                tick label style={font=\footnotesize},
                label style={font=\small},
                legend style={font=\small}
            ]
            \addplot[red, thick] table [x=snr_db, y = mse_db_1] {\data};
            \addplot[blue, thick] table [x=snr_db, y = mse_db_2] {\data};
            \addplot[orange, thick] table [x=snr_db, y = mse_db_3] {\data};
            
            \addplot[black, dash pattern=on 4pt off 2pt, domain=0:20] {-x};
            
            \addplot[red, dash pattern=on 4pt off 2pt, domain=0:20] {10*log10(2*5/(2*10^(x/10))/(32*32*32*32*32))};
            
            \addplot[blue, dash pattern=on 4pt off 2pt, domain=0:20] {10*log10(2*15/(2*10^(x/10))/(32*32*32*32*32))};
            
            \addplot[orange, dash pattern=on 4pt off 2pt, domain=0:20] {10*log10(2*35/(2*10^(x/10))/(32*32*32*32*32))};
            \end{axis}
        \end{tikzpicture}
    }
    \caption{Channel estimation MSE for the various setups. The exact channel model in \eqref{channel} is used. Also shown in dashed are the asymptotes in \eqref{per_entry_crb}.
    The results are averaged over 1000 realizations of the geometric parameters and the noise (100 realizations for the last subfigure). Each subfigure's caption describes the setup as $(\Ntx\times\Nty)\times(\Nrx\times\Nry)\times \Nf$.}
    \label{fig:mse_results_actual}
\end{figure*}
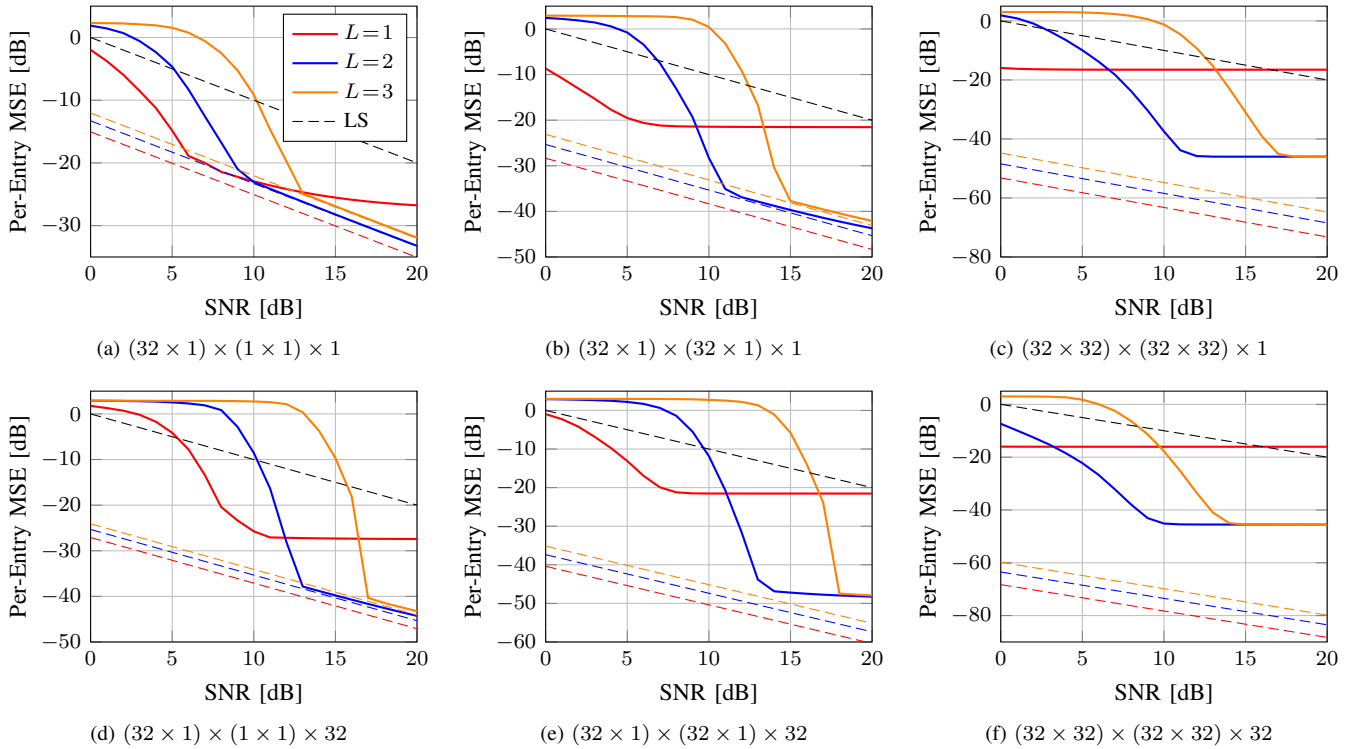


\section{Numerical Evaluation}
\label{sec:numerical_evaluation}

This section assesses the reconstruction MSE of the wavefront parameterization method against the LS estimate, which, for \eqref{observation_model}, amounts to the noisy observation itself with a per-entry MSE of $\frac{1}{\SNR}$.



The three setups in Sec. \ref{sec:infeasibility} are again entertained, at $\fc = 30$~GHz and for either one or multiple frequencies; in the multiple-frequency case, $\df = 5\cdot 10^{-4}$, which corresponds to 15 MHz. Every result is averaged over 1000 realizations of the geometric parameters, $\br$ and $\bR$, uniformly sampled from the spherical shell $\{\br:  5 \text{ m}\leq\|\br\|\leq 15 \text{ m}\}$ and the special orthogonal group, respectively. (The values for $\br$ are purposely small so as to ensure near-field conditions.)

Shown in Fig. \ref{fig:mse_results_unit} are the MSEs for polynomials of degree $L=1$, $2$, and $3$, which determines the phase mismatch.
The amplitude variations over the arrays, which the wavefront estimator ignores, are excluded from the channel model for the figure's top section, but included for the bottom section; the contrast between both sections reveals the significance of the amplitude mismatch.
Also shown is the asymptote in \eqref{per_entry_crb} for each $L$, which bounds what one can hope to achieve with such $L$ and no mismatch in phase or amplitude.

The take-away points from the results are as follows:
\begin{itemize}
    \item The amplitude variations give rise to a high-SNR floor in the UPA-to-UPA setup, but that floor is still tens of dB below the LS performance. The phase-only wavefront estimator is thus applicable to UPAs with $32 \times 32 = 1024$ antennas, but its performance is bound to be compromised if the arrays are substantially larger.
    \item The curves corresponding to $L=1$ describe the performance of a far-field 
    estimator, which is very poor. 
    \item Contingent on the amplitude variations not being significant, the upgrade to $L=2$ renders the performance excellent: beyond an SNR threshold that depends on the setup, the MSE essentially meets its asymptote, indicating that a parabolic wavefront estimator correctly balances accuracy and phase mismatch.
    \item For $L=3$, the estimator suffers from
    an unnecessarily higher SNR threshold, which compromises its otherwise excellent performance.
    \item Availing of observations at multiple frequencies is more beneficial in simpler settings, when there are fewer observations in the spatial domain.
\end{itemize}

Importantly, the SNR threshold exhibited by the wavefront estimator is not fundamental. It can be lowered by sophisticating Alg. \ref{alg:proposed} \cite{mckilliam2014polynomial,do2025multidimensional}
or, without algorithmic refinements, by accruing pilot observations
to enhance the SNR (at the expense of a higher pilot overhead; more on this in the next section).

\section{Summary and Discussion}
\label{sec:conclusion}

The middle way propounded in this paper is ideally suited to estimate near-field LOS channels. It is a parametric method, whereby the number of quantities to estimate is drastically curtailed, yet those quantities are not geometric parameters (deeply hidden from the observations), but rather polynomial coefficients describing the curvature and disposition of the wavefronts (much more accessible from those observations). While the number of polynomial coefficients is slightly higher than the number of geometric parameters, the complexity is radically more graceful in that number. Altogether, wavefront channel estimation yields a superior tradeoff between accuracy and complexity.

Importantly, an antenna-domain estimator such as LS cannot operate with fewer than one pilot symbol per transmit antenna and per frequency, which may represent an insurmountable limitation for large arrays. For the UPA-to-UPA setup in previous sections, for instance, at least $1024$ pilot symbols per frequency are required, highly restricting the applicability.
In contrast, wavefront parametric estimation requires only the handful of pilots specified in the previous section: $(L+1)$ if the transmit array is linear, $(L+1)^2$ if it is planar.
For $L=2$, this amounts to a total of only $9$ pilots in the UPA-to-UPA setup. 
With these $9$ pilots conveyed from a subset of the transmit antennas and frequencies, the reconstruction then implicitly entails an interpolation process for the other (unobserved) transmit antennas and frequencies \cite{do2025near,lu2026near}. This reduction in pilots directly translates to a reduction in overhead. At the same time, a reduction in the total pilot power worsens the estimation SNR and consequently the MSE. As the spectral efficiency is linear in the overhead and sublinear in the MSE \cite[Sec. 4.8]{foundations2018}, shrinking the number of pilots is decidedly beneficial.
Moreover, to the extent that pilot power boosting can be applied, subject to an average power constraint, the benefit of shrinking the number of pilots can be reaped
at no cost in MSE.


As a note of caution, one must always proceed with care when applying parametric estimators because, by their very nature, parametric estimators can only be as good as the models underpinning them \cite{friedlander2019localization, he2021wireless}. If the channel does not abide by the LOS model in \eqref{channel},
then a mismatched estimator reliant on that model will suffer some degradation.
Thus, a first avenue for subsequent research within the confines of LOS channels is to
explicitly account for amplitude variations over the arrays, with a view to accommodating arbitrarily large apertures. A second research direction for LOS channels is to
derive the geometric parameters from the wavefront polynomial coefficients, for the purpose of positioning. 


Transcending LOS channels, the wavefront estimation procedure could also be extended to multipath channels. Advantageously, and contingent only on a sufficiently high polynomial degree,
a wavefront estimator is bound to be transparent to mixtures of near- and far-field multipath components.
The wavefronts for delay-resolvable multipath components could be separately estimated exactly as that of the LOS component. Delay-nonresolvable components, in turn, would call for a multicomponent extension of the multidimensional polynomial phase estimator in \cite{do2025multidimensional}.
%
Thanks to their overparameterization, wavefront estimators are expected to be more robust than their geometric-parameter counterparts, but still somewhat vulnerable to model mismatches, say a presumed number of multipath components that differs from the actual number.
Hence, the extension of wavefront estimators---possibly in conjunction with learning methods---to multipath channels presents a fertile ground for research moving forward.




\appendices

\section{}
\label{app:another_parameterization}

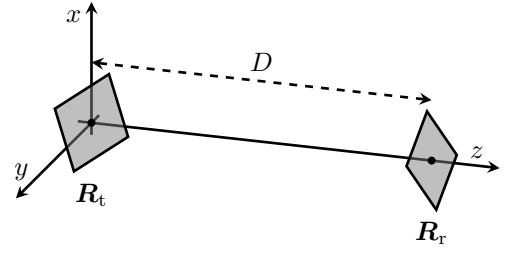
\begin{figure}
    \centering
    \begin{tikzpicture}[
        scale=1, 
        >=stealth, 
        x={(0 cm,0.8 cm)}, 
        y={(-0.5 cm,-0.5 cm)}, 
        z={(0.9 cm,-0.1 cm)}
    ]
        \draw[line width=1pt, ->] 
            (-0.2,0,0) -- (2,0,0) 
            node[above left, yshift = -0.4 cm] {$x$};
        \draw[line width=1pt, ->]
            (0,-0.2,0) -- (0,2,0)
            node[above left, xshift = 0.3 cm, yshift = 0.1 cm] {$y$};
        \draw[line width=1pt, ->]
            (0,0,-0.2) -- (0,0,6)
            node[above, xshift = -0.3 cm] {$z$};
        \draw[line width=1pt, <->, dashed]
            (1,0,0) -- node[midway, above] {$D$} (1,0,5);

        \draw[line width=1pt, fill opacity = 0.5, fill = gray, rotate around x = -15, rotate around y = 20, rotate around z = 0] 
            (0.5,0.5,0) -- (-0.5,0.5,0) -- (-0.5,-0.5,0) -- (0.5,-0.5,0) -- cycle;
        \fill[] (0,0,0) circle (1.5pt);
        \node[] at (-1.2,0.0) {$\bR_{\rm t}$};

        \begin{scope}[shift = {(0,0,5)}]
            \draw[line width=1pt, fill opacity = 0.5, fill = gray, scale = 0.5, rotate around x = 55, rotate around y = -10, rotate around z = 15]
            (1,1,0) -- (-1,1,0) -- (-1,-1,0) -- (1,-1,0) -- cycle;
            \fill[] (0,0,0) circle (1.5pt);
            \node[] at (-1.2,0.0) {$\bR_{\rm r}$};
        \end{scope}
    \end{tikzpicture} 
    \caption{Two planar arrays and the parameters describing their relative geometry. 
    }
    \label{fig:geometry}
\end{figure}

Referring to Fig. \ref{fig:geometry},
the transmit array is in the vicinity of the origin while the receive array is in the vicinity of $D\vec{\bz}$, where $\vec{\bz}\equiv
   [ 0 \;\,  0 \;\,  1 ]^\top$.
The $\nt$th transmit antenna is at
\begin{align}
    \bt_\nt = \bR_{\rm t}
    \begin{bmatrix}
        \dtx \big( \ntx-\frac{\Ntx-1}{2} \big) \\ \dty \big(\nty-\frac{\Nty-1}{2} \big) \\ 0
    \end{bmatrix}
\end{align}
where $\bR_{\rm t}\in\bbR^{3\times 3}$ is a rotation matrix.
Similarly, the locations of the $\nr$th receive antenna can be expressed as
\begin{align}
    \br_\nr = \bR_{\rm r}
    \begin{bmatrix}
        \drx \big( \nrx-\frac{\Nrx-1}{2} \big) \\ \dry \big( \nry-\frac{\Nry-1}{2} \big) \\ 0
    \end{bmatrix}  + D\vec{\bz} ,
\end{align}
where $\bR_{\rm r}\in\bbR^{3\times 3}$ is a rotation matrix.

\begin{table}
\setlength{\tabcolsep}{2pt}
\addtolength{\leftskip} {-2cm}
\addtolength{\rightskip} {-2cm}
\centering
\begin{threeparttable}
\caption{Parameters describing the relative geometry of two arrays}
\label{table:number_of_paramters}
\begin{tabular}{c c c c c c c c c c} 
\toprule
Tx Array & Rx Array & $D$ & $\varphi_{{\rm t},x}$ & $\varphi_{{\rm t},y}$ & $\varphi_{{\rm t},z}$ & $\varphi_{{\rm r},x}$ & $\varphi_{{\rm r},y}$ & $\varphi_{{\rm r},z}$ & \# parameters 
\\
\midrule
linear & single & \cmark & - & \cmark & - & - & - & - & 2 \\
planar & single & \cmark & \cmark & \cmark & - & - & - & - & 3\\
linear & linear & \cmark & - & \cmark & \cmark & - & \cmark & -  & 4\\
planar & linear & \cmark & \cmark & \cmark &  \cmark & - & \cmark & - & 5\\
planar & planar & \cmark & \cmark & \cmark &  \cmark & \cmark & \cmark & - & 6\\
\bottomrule
\end{tabular}
\end{threeparttable}
\end{table}

\begin{figure}
    \captionsetup[subfigure]{format=hang, margin=10pt}
    \centering
    \subfloat[Before rotation.]
    {
    \begin{tikzpicture}[
        scale=1, 
        >=stealth, 
        x={(0 cm,0.8 cm)}, 
        y={(-0.5 cm,-0.5 cm)}, 
        z={(0.9 cm,-0.1 cm)}
    ]
        \draw[line width=1pt, ->] 
            (-0.2,0,0) -- (2,0,0) 
            node[above left, yshift = -0.4 cm] {$x$};
        \draw[line width=1pt, ->]
            (0,-0.2,0) -- (0,2,0)
            node[above left, xshift = 0.3 cm, yshift = 0.1 cm] {$y$};
        \draw[line width=1pt, ->]
            (0,0,-0.2) -- (0,0,6)
            node[above, xshift = -0.3 cm] {$z$};

        \draw[line width=1pt, fill opacity = 0.5, fill = gray] 
            (0.5,0.5,0) -- (-0.5,0.5,0) -- (-0.5,-0.5,0) -- (0.5,-0.5,0) -- cycle;
        \fill[] (0,0,0) circle (1.5pt);
        \node[anchor=west] at (0.1,-0.6) {$\bI$};
        
        \begin{scope}[shift = {(0,0,5)}]
            \draw[line width=1pt]
            (0.5,0,0) -- (-0.5,0,0);
            \fill[] (0,0,0) circle (1.5pt);
            \node[anchor=east] at (-0.3,0.1) {$\bI$};
        \end{scope}
    \end{tikzpicture}
    }\\
    \subfloat[Rotation of the planar array around the $x$ axis, with $\varphi_{{\rm r},x}=0$ from the rotational symmetry of the linear array.]
    {
    \begin{tikzpicture}[
        scale=1, 
        >=stealth, 
        x={(0 cm,0.8 cm)}, 
        y={(-0.5 cm,-0.5 cm)}, 
        z={(1 cm,-0.1 cm)}
    ]
        \draw[line width=1pt, ->] 
            (-0.2,0,0) -- (2,0,0) 
            node[above left, yshift = -0.4 cm] {$x$};
        \draw[line width=1pt, ->]
            (0,-0.2,0) -- (0,2,0)
            node[above left, xshift = 0.3 cm, yshift = 0.1 cm] {$y$};
        \draw[line width=1pt, ->]
            (0,0,-0.2) -- (0,0,6)
            node[above, xshift = -0.3 cm] {$z$};

        \draw[line width=1pt, fill opacity = 0.5, fill = gray, rotate around x = -15] 
            (0.5,0.5,0) -- (-0.5,0.5,0) -- (-0.5,-0.5,0) -- (0.5,-0.5,0) -- cycle;
        \fill[] (0,0,0) circle (1.5pt);
        \node[anchor=west] at (0.1,-0.6) {$\bR_x(\varphi_{{\rm t},x})$};
        
        \begin{scope}[shift = {(0,0,5)}]
            \draw[line width=1pt]
            (0.5,0,0) -- (-0.5,0,0);
            \fill[] (0,0,0) circle (1.5pt);
            \node[anchor=east] at (-0.3,0.1) {$\bR_x(0)$};
        \end{scope}
    \end{tikzpicture}
    }\\
    \subfloat[Rotation of each array around the $y$ axis.]
    {
    \begin{tikzpicture}[
        scale=1, 
        >=stealth, 
        x={(0 cm,0.8 cm)}, 
        y={(-0.5 cm,-0.5 cm)}, 
        z={(1 cm,-0.1 cm)}
    ]
        \draw[line width=1pt, ->] 
            (-0.2,0,0) -- (2,0,0) 
            node[above left, yshift = -0.4 cm] {$x$};
        \draw[line width=1pt, ->]
            (0,-0.2,0) -- (0,2,0)
            node[above left, xshift = 0.3 cm, yshift = 0.1 cm] {$y$};
        \draw[line width=1pt, ->]
            (0,0,-0.2) -- (0,0,6)
            node[above, xshift = -0.3 cm] {$z$};

        \draw[line width=1pt, fill opacity = 0.5, fill = gray, rotate around x = -15, rotate around y = 20] 
            (0.5,0.5,0) -- (-0.5,0.5,0) -- (-0.5,-0.5,0) -- (0.5,-0.5,0) -- cycle;
        \fill[] (0,0,0) circle (1.5pt);
        \node[anchor=west] at (0.1,-0.6) {$\bR_y(\varphi_{{\rm t},y})\bR_x(\varphi_{{\rm t},x})$};

        \begin{scope}[shift = {(0,0,5)}]
            \draw[line width=1pt, rotate around y = -10]
            (0.5,0,0) -- (-0.5,0,0);
            \fill[] (0,0,0) circle (1.5pt);
            \node[anchor=east] at (-0.3,0.1) {$\bR_y(\varphi_{{\rm r},y})\bR_x(0)$};
        \end{scope}
    \end{tikzpicture}
    }\\
    \subfloat[Rotation of the planar array around $z$, with $\varphi_{{\rm r},z}=0$ for the linear array as the relative geometry is invariant under the substitution $(\varphi_{{\rm t},z},\varphi_{{\rm r},z}) \gets (\varphi_{{\rm t},z}-\varphi_{{\rm r},z},0)$. ]
    {
    \begin{tikzpicture}[
        scale=1, 
        >=stealth, 
        x={(0 cm,0.8 cm)}, 
        y={(-0.5 cm,-0.5 cm)}, 
        z={(1 cm,-0.1 cm)}
    ]
        \draw[line width=1pt, ->] 
            (-0.2,0,0) -- (2,0,0) 
            node[above left, yshift = -0.4 cm] {$x$};
        \draw[line width=1pt, ->]
            (0,-0.2,0) -- (0,2,0)
            node[above left, xshift = 0.3 cm, yshift = 0.1 cm] {$y$};
        \draw[line width=1pt, ->]
            (0,0,-0.2) -- (0,0,6)
            node[above, xshift = -0.3 cm] {$z$};

        \draw[line width=1pt, fill opacity = 0.5, fill = gray, rotate around x = -15, rotate around y = 20, rotate around z = -15] 
            (0.5,0.5,0) -- (-0.5,0.5,0) -- (-0.5,-0.5,0) -- (0.5,-0.5,0) -- cycle;
        \fill[] (0,0,0) circle (1.5pt);
        \node[anchor=west] at (0.1,-0.6) {$\bR_z(\varphi_{{\rm t},z}-\varphi_{{\rm r},z})\bR_y(\varphi_{{\rm t},y})\bR_x(\varphi_{{\rm t},x})$};
        
        \begin{scope}[shift = {(0,0,5)}]
            \draw[line width=1pt, rotate around x = 55, rotate around y = -10]
            (0.5,0,0) -- (-0.5,0,0);
            \fill[] (0,0,0) circle (1.5pt);
            \node[anchor=east] at (-0.3,0.1) {$\bR_z(0)\bR_y(\varphi_{{\rm r},y})\bR_x(0)$};
        \end{scope}
    \end{tikzpicture}
    }
    \caption{Sufficiency of five parameters to describe the relative geometry of a planar and a linear array.}
    \label{fig:five_parameters_suffice}
\end{figure}
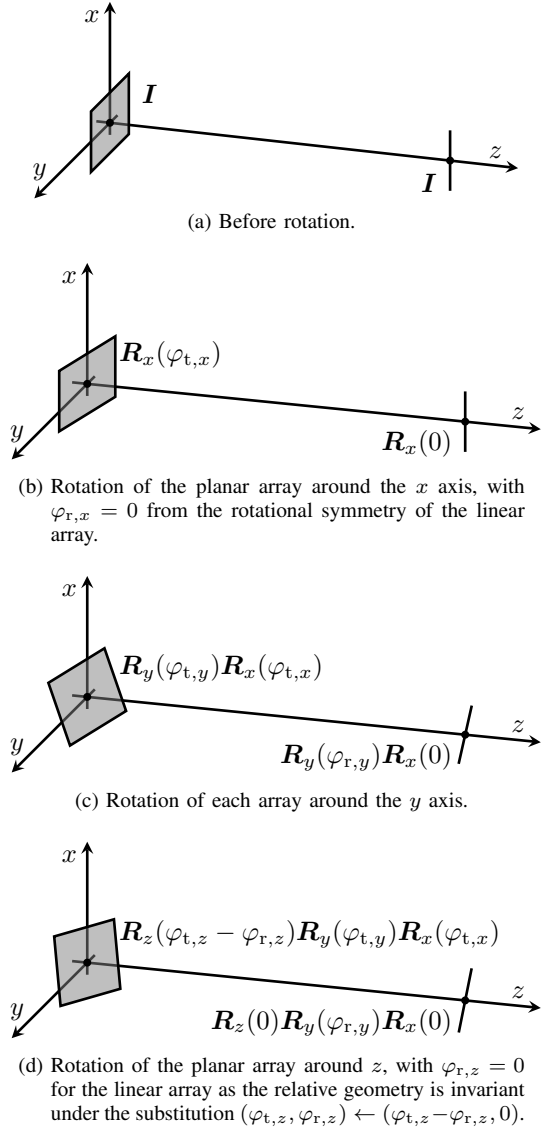



Both $\bR_{\rm t}$ and $\bR_{\rm r}$ can be posed as a composition of matrices rotating with respect to the $x$, $y$, and $z$ axes \cite[Ch. 8.3]{dunn20113d}, i.e.,
\begin{align}
    &\bR_{\rm t} = \bR_{z}(\varphi_{{\rm t},z})\bR_{y}(\varphi_{{\rm t},y})\bR_{x}(\varphi_{{\rm t},x})\\ &\bR_{\rm r} = \bR_{z}(\varphi_{{\rm r},z})\bR_{y}(\varphi_{{\rm r},y})\bR_{x}(\varphi_{{\rm r},x})
\end{align}
where the angles lie on $[-\pi,\pi)$ and 
\begin{align}
    \bR_x(\varphi) & = \begin{bmatrix}
        1 & 0 & 0\\
        0 & \cos\varphi & -\sin\varphi\\
        0 & \sin\varphi & \cos\varphi
    \end{bmatrix} \nonumber  \\ \label{rotationMatrices}
    \bR_y(\varphi) & = \begin{bmatrix}
        \cos\varphi & 0 & \sin\varphi\\
        0 & 1 & 0\\
        -\sin\varphi & 0  & \cos\varphi
    \end{bmatrix} \\ \nonumber
    \bR_z(\varphi) & = \begin{bmatrix}
        \cos\varphi & -\sin\varphi & 0\\
        \sin\varphi & \cos\varphi & 0\\
        0 & 0 & 1
    \end{bmatrix}.
\end{align}

The relative position and orientation of the receive array from the vantage of the transmit array is fully captured by $\bR_{\rm t}$, $\bR_{\rm r}$, and $D$. Since each of these rotation matrices can be described by three angles, the total number of geometric parameters is seven. As there are only six
degrees of freedom---three for rotation and three for translation---describing any rigid motion of a single body, the foregoing parameterization is redundant. Without loss of generality, one can let $\varphi_{{\rm r},z} = 0$. As presented in Table \ref{table:number_of_paramters}, one can similarly set some angles to zero without loss of generality. For instance, the sufficiency of five parameters for a combination of planar and linear arrays is illustrated in Fig. \ref{fig:five_parameters_suffice}. 



\begin{figure*}
    \centering
    \subfloat[Considered geometry\label{fig:considered_geometry}]
    {
        \begin{tikzpicture}[>=stealth]
            \clip (-0.5,-2.4) rectangle (7,2.4);
            
            \draw[line width=1pt, ->] (-0.5,0) -- (7,0) node[below left, xshift = -0.05, yshift = -0.05 cm] {$z$};
            \draw[line width=1pt, ->] (0,-2) -- (0,2) node[below left, xshift = -0.05, yshift = -0.05 cm] {$x$};
            
            \draw[line width=3pt] (0,-0.4) -- (0,0.4);
            
            \node[draw,circle,fill=black,scale=0.3] at (6,0) [] {};

            \draw[line width=1pt, <->, dashed] (0,0.8) -- node[midway, above] {$D$} (6,0.8);

        \end{tikzpicture} 
    }
    \subfloat[Objectives as a function of $D'$\label{fig:wiggle} for $D= 5$ m]
    {
            \begin{tikzpicture}
            \pgfplotstableread[col sep=comma]{data/cost_landscape.csv}{\data}
            \begin{axis}[
                width=8.0cm, height=4.8cm,
                xlabel={$D'$},
                ylabel={Square-root of objective},
                xlabel style={yshift=1mm},
                ylabel style={yshift=-3mm},
                xtick={4.6, 4.8, ..., 5.4},
                xmin=4.6, xmax=5.4,
                ymin=0, ymax=40,
                ytick={0, 10, ..., 40},
                grid=none,
                legend pos=north east,
                legend cell align={left},
                tick label style={font=\footnotesize},
                label style={font=\small},
                legend style={font=\small}
            ]
            \addplot[black] table [x=z, y = point] {\data};
            \addlegendentry{\eqref{mle}}
            \addplot[black, line width = 2pt] table [x=z, y = plane] {\data};
            \addlegendentry{\eqref{mle_cauchy}}
            \end{axis}
        \end{tikzpicture}
    }
    \caption{Comparison between objectives \eqref{mle} and \eqref{mle_cauchy} in an exemplary setting.}
    \label{fig:working_example}
\end{figure*}
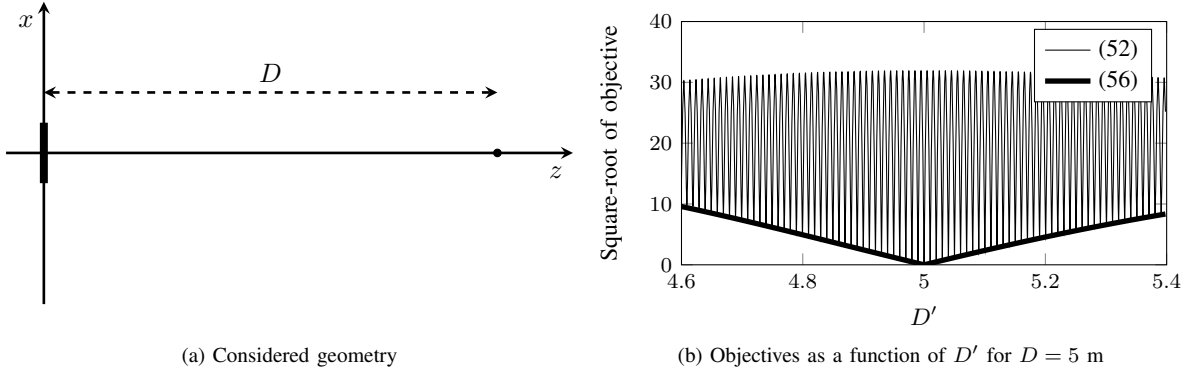

\section{}
\label{app:implementation_detail}

This appendix sets forth a numerical method to compute the MLE of $\br$ and $\bR$. For numerical efficiency (see App.~\ref{app:mle}),
rather than
$\frac{1}{|[\bN]|}  \sum_{\bn} |y(\bn) - h(\bn;\br,\bR)|^2$,
the cost function is
\begin{align}
    \sum_{\bn} |y(\bn)|^2  - \frac{\Big|\sum\limits_{\bn} y(\bn) \overline{h(\bn;\br,\bR)}\Big|^2}{\sum\limits_{\bn} |h(\bn;\br,\bR)|^2} ,
\end{align}
normalized by $|[\bN]|$.

Since $\bR$ is a rotation matrix, direct application of gradient-based methods requires a parameterization. As any rotation matrix can be expressed as the matrix exponential of a skew-symmetric matrix, parameterized by three values, for an initialization $\bR_0$ one can parameterize $\bR$ by $\bomega = (\omega_1, \omega_2, \omega_3)$ as
\begin{align}
    \bR = \bR_0 \exp\left(\begin{bmatrix}
        0 & -\omega_3 & \omega_2\\
        \omega_3 & 0 & -\omega_1\\
        -\omega_2 & \omega_1 & 0\\
    \end{bmatrix}\right).
\end{align}
To optimize $\br$ and $\bomega$, an Adam optimizer with learning rate $0.01$ is employed, with other hyperparameters set to the PyTorch default values and with $2^{10}$ initializations of $\br$ and $\bR$, uniformly sampled from the spherical shell $\{\br:  5 \text{ m}\leq\|\br\|\leq 15 \text{ m}\}$ and the special orthogonal group.

\section{}
\label{app:mle}

On the basis of the observation model in \eqref{observation_model}, the MLE is 
\begin{align}
    \argmin_{\btheta} \sum_{\bn} |y(\bn) - h(\bn;\btheta)|^2. \label{mle}
\end{align}
Alternatively, by incorporating a complex attenuation $\beta \in\bbC$ such that
    \begin{align}
        y(\bn) = \beta h(\bn;\btheta) + w_\bbC(\bn), \label{observation_model_with_attenuation}
    \end{align}
the MLE becomes
\begin{align}
  &\argmin_{\beta, \btheta} \sum_{\bn}  |y(\bn) - \beta h(\bn;\btheta)|^2 .
  \label{mle_attenuation}
\end{align}
For given $\btheta$, the above is a closest-point problem \cite[Eq. 4.7]{steele2004cauchy}, which is an instance of the LS problem. Optimizing over $\beta$ gives
\begin{align}
    \hat{\beta} = \frac{\sum_\bn y(\bn)\overline{h(\bn;\btheta)}}{\sum_\bn |h(\bn;\btheta)|^2},    
\end{align}
whereby \eqref{mle_attenuation} becomes
\begin{align}
    &\argmin_{\btheta}\!\!\left[\sum_{\bn} |y(\bn)|^2 \! -\! \frac{\Big|\!\sum\limits_{\bn} y(\bn) \overline{h(\bn;\btheta)}\Big|^2}{\sum\limits_{\bn} |h(\bn;\btheta)|^2}\right], \label{mle_cauchy} 
\end{align}
with $\overline{\cdot}$ denoting complex conjugation. 

Instead of optimizing $\beta$ over the complex plane, one can constrain it to the unit circle, in which case \cite{rife1974single}
\begin{align}
    \hat{\beta} = \frac{\sum_\bn y(\bn)\overline{h(\bn;\btheta)}}{\big|\sum_\bn y(\bn)\overline{h(\bn;\btheta)}\big|},    
\end{align}
turning \eqref{mle_attenuation} into
\begin{align}
    \argmin_{\btheta}\!\!\left[\sum_{\bn} \big(|y(\bn)|^2+ |h(\bn;\btheta)|^2\big) \! -\! 2\bigg|\!\sum\limits_{\bn} y(\bn) \overline{h(\bn;\btheta)}\bigg|\right]. \label{mle_am_gm}
\end{align}

Provided that the channel has a constant amplitude,
\eqref{mle} reduces to
\begin{align}
    \argmax_{\btheta}\Re \bigg\{\sum\limits_{\bn} y(\bn) \overline{h(\bn;\btheta)}\bigg\}
\end{align}
while both \eqref{mle_cauchy} and \eqref{mle_am_gm} boil down to
\begin{align}
    \argmax_{\btheta}\bigg|\sum\limits_{\bn} y(\bn) \overline{h(\bn;\btheta)}\bigg|. \label{mlePeriodogram}
\end{align}

Armed with a parameter estimate $\hat{\btheta}$, one can then reconstruct the channel as 
\begin{align}
    \hat{h}(\bn) \equiv \hat{\beta} h(\bn;\hat{\btheta}),
\end{align}
where
\begin{align}
    \hat{\beta} = \begin{cases}
        1 &\text{for }\eqref{mle}\\
        \frac{\sum_\bn y(\bn)\overline{h(\bn;\hat{\btheta})}}{\sum_\bn |h(\bn;\hat{\btheta})|^2} & \text{for }\eqref{mle_cauchy}\\
        \frac{\sum_\bn y(\bn)\overline{h(\bn;\hat{\btheta})}}{|\sum_\bn y(\bn)\overline{h(\bn;\hat{\btheta})}|} & \text{for }\eqref{mle_am_gm}
    \end{cases}.
\end{align}

The explicit incorporation of the attenuation $\beta$ has two effects:
\begin{itemize}
    \item The objective function exhibits a more favorable behavior with respect to the communication distance, in the vicinity of the global optimum.
    \item The parameter count grows by one and by two, respectively, depending on whether $\beta$ is on the unit circle or the complex plane. The estimation error worsens, but only by a fraction of dB. 
\end{itemize}

Numerically solving \eqref{mle}, \eqref{mle_cauchy}, or \eqref{mle_am_gm} entails a coarse search over a grid---some works invoke the fast Fourier transform to reduce the computational burden, which necessitates some approximations---followed by a fine search; for the coarse search, one may resort to a random search instead of the grid search, i.e., using multiple random initializations.
The number of grid points grows exponentially with the number of parameters, rendering the MLE computationally unwieldy for four and especially for five or six geometric parameters, say for
transmit and receive planar arrays (recall Sec. \ref{sec:geometric_parameterization}).

From a computational standpoint, \eqref{mle_cauchy} and \eqref{mle_am_gm}
are preferable to
\eqref{mle},
as the latter can wiggle wildly even without noise and more grid points are therefore required.
To exemplify this phenomenon, let us consider the noiseless connection between a 256-antenna ULA transmitter and a single-antenna receiver with half-wavelength spacing (see Fig. \ref{fig:considered_geometry}) at 30 GHz. The receiver is on the $z$-axis, whereby $\br = D\vec{z}$ and
\begin{align}
    h(\ntx; D) &= \frac{D}{\sqrt{D^2+ d_{{\mathrm t},x}^2\big(\ntx-\tfrac{\Ntx-1}{2}\big)^2}}\\
    &\quad \cdot\exp\!\bigg(\!-j\frac{2\pi}{\lambdac} \sqrt{D^2+ d_{{\mathrm t},x}^2\big(\ntx-\tfrac{\Ntx-1}{2}\big)^2}\bigg). \nonumber
\end{align}

In the absence of noise, the observation $y$ at some distance is $h$ itself.
If $D$ is the true distance giving rise to the observation and $D'$ is some other distance being considered to parameterize the estimate, the objectives in \eqref{mle} and \eqref{mle_cauchy} are contrasted in Fig.~\ref{fig:wiggle}. More precisely, and for a better visualization, the square-root of each objective is displayed.


For a small distance deviation $\Delta$, 
\begin{align}
    h(\ntx;D'+\Delta) \approx e^{-j\frac{2\pi}{\lambdac}\Delta} h(\ntx;D'),
\end{align}
whereby \eqref{mle} becomes 
\begin{align}
    &\sum_{\ntx} \left( |h(\ntx;D)|^2 + |h(\ntx;D'+\Delta)|^2 \right) \\
    &\qquad\qquad +  2\Re\bigg\{\sum\limits_{\ntx} h(\ntx;D) \overline{h(\ntx;D'+\Delta)}\bigg\} \nonumber \\
    &\quad \approx \sum_{\ntx} \left( |h(\ntx;D)|^2 + |h(\ntx;D'+\Delta)|^2 \right) \\
    &\qquad\qquad   +2\Re\bigg\{e^{j\frac{2\pi}{\lambdac}\Delta}\sum_{\ntx} h(\ntx;D)\overline{h(\ntx;D')}\bigg\}, \nonumber
\end{align}
where the wiggling behavior arises on account of $e^{j\frac{2\pi}{\lambdac}\Delta}$. This term actually does not depend on the channel, but it merely reflects the carrier oscillations, and the introduction of a complex attenuation enables disassociating the tracking of those oscillations from the estimation of the baseband channel; for that, the phase of $\beta$ suffices. If $\beta$ is allowed to live outside the unit circle, then it can further absorb certain aspects of the baseband channel.

\section{}
\label{app:legendre}

The ultraspherical polynomials, $C_\ell^\alpha(x)$, are defined via the power series expansion \cite[Eq. 22.9.3]{abramowitz1948handbook}
\begin{align}
    \frac{1}{(1 + 2xt + t^2)^{\alpha}} = \sum_{\ell=0}^\infty C_\ell^\alpha(x) \, (-t)^\ell. \label{ultraspherical_definition}
\end{align}
In the case at hand, the interest is in the series expansion of $\sqrt{1 + 2xt + t^2}$, and therefore in $C_\ell^{-1/2}(x)$. The recurrence relation \cite[Eq. 22.7.23]{abramowitz1948handbook}
\begin{align}
    C_{\ell}^{-\frac{1}{2}}(x) &= \frac{1}{2\ell-1} \! \left(C_{\ell-2}^{\frac{1}{2}}(x)-C_{\ell}^{\frac{1}{2}}(x)\right)
\end{align}
recasts the problem as that of computing $C_\ell^{1/2}(x)$, which equals the Legendre polynomial $P_\ell(x)$ \cite[Eq. 22.5.36]{abramowitz1948handbook}. The initial Legendre polynomials are 
\begin{align}
P_0(x) & = 1 \\
P_1(x) & = x \\
P_2(x) & = {\textstyle \frac{1}{2}}(3x^2-1) \\
P_3(x) & = {\textstyle \frac{1}{2}}(5x^3-3x) \\
P_4(x) & = {\textstyle \frac{1}{8}}(35x^4-30x^2+3) \\
P_5(x) & = {\textstyle \frac{1}{8}}(63x^5-70x^3+15x) .
\end{align}
Combining the above yields the expansion
\begin{align}
    &\sqrt{1+2xt + t^2} \nonumber\\
    &\qquad = 1+xt + \sum_{\ell=2}^\infty \frac{1}{2\ell-1} \big(P_{\ell-2}(x)-P_\ell(x)\big)(-t)^\ell\\
    & \qquad = 1+xt + \frac{1}{2}(1-x^2)t^2 - \frac{1}{2}(x-x^3)t^3\\
    &\qquad \quad - \frac{1}{8}(1-6x^2+5x^4)t^4 + \cdots. \nonumber
\end{align}

\bibliographystyle{IEEEtran}
\bibliography{ref}

\end{document}